\documentclass[nojss]{jss}

\usepackage{orcidlink,thumbpdf,lmodern}

\usepackage[utf8]{inputenc}

\author{
Łukasz Chrostowski\\Analyx \And Piotr
Chlebicki~\orcidlink{0009-0006-4867-7434}\\Stockholm University
\AND Maciej Beręsewicz~\orcidlink{0000-0002-8281-4301}\\Poznań
University of Economics and Business\\
Statistical Office in Poznań
}
\title{\pkg{nonprobsvy} -- An R package for modern methods for
non-probability surveys}

\Plainauthor{Łukasz Chrostowski, Piotr Chlebicki, Maciej Beręsewicz}
\Plaintitle{nonprobsvy -- An R package for modern methods for
non-probability surveys}
\Shorttitle{\pkg{nonprobsvy} for non-probability surveys}

\Abstract{
The paper presents \pkg{nonprobsvy} -- an \proglang{R} package for
inference based on non-probability samples. The package implements
various approaches that can be categorized into three groups:
prediction-based approach, inverse probability weighting and doubly
robust approach. In the package, we assume the existence of either
population-level data or probability-based population information and
leverage the \pkg{survey} package for inference. The package implements
both analytical and bootstrap variance estimation for the proposed
estimators. In the paper we present the theory behind the package, its
functionalities and a~case study that showcases the usage of the
package. The package is aimed at scientists and researchers who would
like to use non-probability samples (e.g., big data, opt-in web panels,
social media) to accurately estimate population characteristics.
}

\Plainkeywords{data integration, doubly robust estimation, propensity
score estimation, mass imputation, survey}

\Address{
    Łukasz Chrostowski\\
    Analyx\\
    Krysiewicza 2\\
61-887 Poznań\\
  E-mail: \email{lukaszchrostowski6@gmail.com}\\
  
      Piotr Chlebicki\\
    Stockholm University\\
    Matematiska institutionen\\
Albano hus 1\\
106 91 Stockholm, Sweden\\
  E-mail: \email{piotr.chlebicki@math.su.se}\\
  URL: \url{https://www.su.se/profiles/pich3772}\\~\\
      Maciej Beręsewicz\\
    Poznań University of Economics and Business\\
Statistical Office in Poznań\\
    \hfill\break
Poznań University of Economics and Business\\
Department of Statistics\\
Institute of Informatics and Quantitative Economics\\
Al. Niepodległosci 10\\
61-875 Poznań, Poland\\
\strut \\
Centre for the Methodology of Population Studies\\
Statistical Office in Poznań\\
ul. Wojska Polskiego 27/29\\
60-624 Poznań, Poland\\
  E-mail: \email{maciej.beresewicz@ue.poznan.pl}\\
  URL: \url{https://maciejberesewicz.com}\\~\\
  }

\usepackage{amsmath, amsthm, amssymb} \usepackage{calc, ragged2e} \usepackage[ruled]{algorithm2e} \usepackage{algpseudocode}                                   

\begin{document}

\section{Introduction}\label{sec-introduction}

In official statistics, information about the target population and its
characteristics is mainly collected through probability surveys,
censuses or is obtained from administrative registers and covers all (or
nearly all) units of the population. However, owing to increasing
non-response rates, particularly unit non-response and non-contact,
which result from the growing respondent burden as well as rising costs
of surveys conducted by National Statistical Institutes, non-probability
data sources are becoming more popular
\citep{berkesewicz2017two, beaumont2020probability, biffignandi2021handbook}.
Non-probability surveys, such as opt-in web panels, social media,
scanner data, mobile phone data or voluntary register data, are
currently being explored for use in the production of official
statistics \citep{citro2014multiple, daas2015big}, public opinion
studies \citep{Schonlau2017} or market research \citep[cf.][]{Grow2022}.
Since the selection mechanism underlying these sources is unknown,
standard design-based inference methods cannot be directly applied and,
in the case of large datasets, can lead to the big data paradox, i.e.,
the larger the sample, the larger the bias, as described by
\citet{meng2018statistical}.

Table \ref{tab-comparison-characteristics} compares basic
characteristics of probability and non-probability samples. In
particular, it shows the advantages and disadvantages of each type with
respect to the selection mechanism, the population coverage, bias,
variance, costs and timeliness. In general, the quality of
non-probability samples suffers from an unknown selection mechanism
(i.e., unknown probabilities of inclusion) and under-coverage of certain
groups from the population (e.g., older people). As a~result, direct
estimates based on non-probability samples are biased and, in most
cases, are characterized by small variance owing to their size.
Certainly, the costs and timeliness of these surveys are significantly
smaller than those of probability samples.

\begin{table}[ht!]
    \centering
    \begin{tabular}{lll}
    \hline
    \textbf{Factor}   &  \textbf{Probability sample} & \textbf{Non-probability sample}\\
    \hline
    Selection & Known probabilities & Unknown probabilities (self-selection) \\
    Coverage & Complete & May be incomplete \\
    Estimation bias & Unbiased under design & Potential systematic bias \\
    Variance of estimates & Typically high & Typically low \\
    Cost & High & Low \\
    Timeliness & Long delay & Very short delay \\
    \hline
    \end{tabular}
    \caption{A comparison of probability and non-probability samples and their characteristics.}
    \label{tab-comparison-characteristics}
\end{table}

To address this problem, several approaches have been proposed, which
rely on the estimation of propensity scores (PS; i.e., inclusion
probabilities) for deriving inverse probability weights (IPW; also known
as propensity score weighting/adjustment,
cf.~\citet{lee2006propensity, lee2009estimation}), on model-based
prediction (in particular, mass imputation estimators; MI) and on the
doubly robust (DR) approach involving IPW and MI estimators. Two main
scenarios are usually considered:

\begin{enumerate}
\item only population-level means or totals are available (e.g., from census, registers or sample surveys), 
\item unit-level data are available either in the form of registers covering the whole population or in the form of probability surveys \citep[cf.][]{elliott_inference_2017}. 
\end{enumerate}

\citet{wu2022statistical} classified these approaches into three groups
that require a~joint randomization framework involving a~probability
sampling design (denoted as \(p\)) and an outcome regression model
(denoted as \(\xi\)) or a~PS model (denoted as \(q\)). According to this
classification, IPW estimators represent the \(qp\) framework, MI
estimators represent the \(\xi p\) framework, and DR estimators can
represent either the \(qp\) or the \(\xi p\) framework.

Most approaches assume that population data is available in order to
reduce the bias of non-probability sampling by reweighting to reproduce
known population totals/means (i.e., IPW estimators); by modeling the
target variable using various techniques (i.e., MI estimators); or by
combining both approaches (e.g., DR estimators,
cf.~\citet{chen2020doubly}; see also multilevel regression and
post-stratification, MRP; Mister-P,
cf.~\citet{gelman1997poststratification}). This topic has become very
popular and a~number of new methods have been proposed; for instance
non-parametric approaches based on nearest neighbours
\citep{yang2021integration}, kernel density estimation
\citep{chen_nonparametric_2022}, empirical likelihood
\citep{kim2023empirical}, model-calibration with LASSO \citep{chen2018}
or quantile balanced IPW \citep{beresewicz2025} to name a~few. It should
be highlighted that, in contrast to probability samples, there is no
single method that can be used for all non-probability samples. Based on
the methods available in the literature several statistical software
solutions have been developed. These pieces of software will be
presented and compared with \pkg{nonprobsvy} \citep{nonprobsvy} in the
next section.

\subsection{Software for non-probability samples}\label{sec-software}

Table \ref{tab-comparisons} presents a~comparison of selected packages
in terms of the availability of different inference methods. The reason
for this is the availability of software designed for inference based on
non-probability samples, not a~general way to estimate propensity scores
or model \(E(Y|X)\). For such reviews, one can refer to the
\citet[Chapter 6]{valliant2018survey} who discussed weighting and
matching in this context in \proglang{Stata} \citep{Stata2025} using
\code{logit}, \code{svy} and \code{teffects nnmatch} commands or
\citet[e.g., Chapter 8]{lohr2021sas} who covers \proglang{SAS} software
(\citet{sas94}; e.g., \code{PROC LOGISTIC} procedure).

We focus on packages available as open source or via CRAN or PyPI (for
non-CRAN or non-PyPI software see \citet{cobo2024software}). Our
comparison includes five packages that focus specifically on
non-probability sampling: in \proglang{R} -- \pkg{NonProbEst}
\citep{NonProbEst} and our \pkg{nonprobsvy}, and in \proglang{Python} --
\pkg{balance} \citep{sarig2023balancepythonpackage}, \pkg{inps}
\citep{castro2024inps}, and in \proglang{Stan} \citep{carpenter2017stan}
e.g., through the \proglang{R} package \pkg{rstanarm}, which allows you
to easily apply the \pkg{rstanarm}'s MRP method \citep{rstanarm}.

\begin{table}[ht!]
\centering
\begin{tabular}{lccccc}
\hline
 & & \multicolumn{2}{c}{\proglang{Python}} & \multicolumn{2}{c}{\proglang{R}} \\
\cline{3-4} \cline{5-6}
\textbf{Functionalities} & \proglang{Stan} & \pkg{balance} & \pkg{inps} & \pkg{NonProbEst} & \pkg{nonprobsvy} \\
\hline
IPW  & -- & $\checkmark$ & $\checkmark$ & $\checkmark$ & $\checkmark$ \\
Calibrated IPW  & -- & -- & -- & -- &$\checkmark$ \\
MI & -- & -- & -- & $\checkmark$ & $\checkmark$ \\
DR & -- & -- & $\checkmark$ & -- & $\checkmark$\\
MRP & $\checkmark$ & -- & -- & -- & --\\
Variable selection & $\checkmark$ & $\checkmark$ & $\checkmark$ & $\checkmark$ & $\checkmark$\\
Analytical variance & -- & -- & -- & -- &  $\checkmark$\\
Bootstrap variance & -- & -- & -- & $\checkmark$ & $\checkmark$\\
Integration with \pkg{survey} or \pkg{samplics} & -- & -- & -- & -- & $\checkmark$\\
\hline
\end{tabular}
\caption{A comparison of packages and implemented methods.}
\label{tab-comparisons}
\end{table}

The \pkg{NonProbEst} package, while more limited than \pkg{nonprobsvy},
offers various techniques, such as PS or prediction approaches (e.g.,
model-calibrated). Users can choose several different settings for PS
weights, variable selection and can estimate variance of the mean using
the leave-one-out Jackknife procedure. Unfortunately, the package is no
longer actively developed (as of August 2025) and some of the techniques
are either outdated or have been shown to be inappropriate for
non-probability samples. While the package contains functions designed
for specific methods, it does not allow users to leverage the
\pkg{survey} package \citep{survey-pkg} for estimation. The
\pkg{balance} package is solely dedicated to the PS approach (as of
version 0.10.0). It assumes that the reference probability sample is
available and the authors have implemented the variance estimator of the
weighted mean as a~measure of uncertainty for the IPW estimator. The
weights for the IPW estimator are constructed using the approach
proposed by \citet{Schonlau2017}. The \pkg{inps} package supports the
use of unit-level data from a~probability sample or the population,
implements IPW, MI and DR estimators, and offers users the possibility
of selecting variables but is still at a~very early stage of
development. It also implements kernel weighting and a~simple bootstrap
approach via the \pkg{scipy.stats} package \citep{scipy2020}. Neither
\pkg{balance} nor \pkg{inps} supports the use of the \pkg{samplics}
package \citep{Diallo2021}. Finally, we note that the MRP approach is
implemented solely in \proglang{Stan} with variable selection specified
by an appropriate prior, but it can be easily implemented in other
probabilistic programming languages, e.g., using the \pkg{Turing.jl}
package \citep{turing} in the \proglang{Julia} language \citep{julia}.

The \pkg{nonprobsvy} package has several advantages over those presented
above. Firstly, it implements state-of-the-art methods recently proposed
in the literature, along with valid statistical inference procedures.
Secondly, it offers other approaches, such as calibrated IPW (where PS
weights match either the population or estimated totals), NN and PMM
matching, various IPW and DR estimators using generalized linear models
(GLM) with different link functions for probability estimation. Thirdly,
it supports the functions included in the \pkg{survey} package to
account for the design of the probability sample. Finally, we provide a
user-friendly API that mimics \code{glm()}, \code{survey::calibrate()}
and other functions known in \proglang{R}, together with the main
function to specify the approach and estimators. As far as we know, the
\pkg{nonprobsvy} is the only software (open-access or commercial) that
offers such functionalities.

The remaining part of the paper is structured as follows. Section
\ref{sec-methods} is dedicated to the theory of statistical inference
based on non-probability samples. We provide the basic set-up and
introduce specific methods in separate sections. We follow the notation
used by \citet{wu2022statistical} throughout the paper. Section
\ref{sec-package} describes the main function and the package
functionalities. Section \ref{sec-data-analysis} presents an empirical
study showcasing the process of integrating data from the Polish Job
Vacancy Survey with voluntary administrative data from the Central Job
Offers Database in order to estimate the number of companies with at
least one vacancy offered on a~single shift. Section \ref{sec-s3methods}
presents classes and \code{S3} methods implemented in the package. The
paper ends with a~summary and plans for future development. The Appendix
contains Section \ref{sec-details}, which presents algorithms for
selected MI estimators.

\section{Methods for non-probability samples}\label{sec-methods}

\subsection{Basic setup}\label{basic-setup}

Let \(U=\{1,..., N\}\) denote the target population consisting of \(N\)
labeled units. Each unit \(i\) has an associated \(L\)-dimensional
vector of auxiliary variables \(\boldsymbol{x}_{i}\) and the study
(target) variable \(y_{i}\). Let
\(\{ (y_i, \boldsymbol{x}_i), i \in \mathcal{I}_{\text{NP}}\}\) be a
non-probability sample \(S_{\text{NP}}\) of size \(n_{\text{NP}}\),
where \(\mathcal{I}_{\text{NP}}\) is an index set for \(S_{\text{NP}}\).
Let
\(\left\{\left(\boldsymbol{x}_i, \pi_{\text{P},i}\right), i \in \mathcal{I}_{\text{P}}\right\}\)
be a~probability sample \(S_{\text{P}}\) of size \(n_{\text{P}}\), where
\(\mathcal{I}_{\text{P}}\) is an index set for \(S_{\text{P}}\), and the
only information known for all units in the population refers to
auxiliary variables \(\boldsymbol{x}\) and inclusion probabilities
\(\pi_{\text{P}}\). Each unit in the \(S_{\text{P}}\) sample has been
assigned a~design-based weight given by
\(d_{\text{P},i} = 1/\pi_{\text{P},i}\).

Let \(I_{\text{NP},i}=I(i \in S_{\text{NP}})\) and
\(I_{\text{P},i}=I(i \in S_{\text{P}})\) be indicators of inclusion in
the non-probability sample \(S_{\text{NP}}\) and the probability sample
\(S_{\text{P}}\), respectively, which are defined for all units in the
target population. Let
\(\pi_{\text{NP},i}=P(I_{\text{NP},i}=1 \mid \boldsymbol{x}_i, y_i)=P(I_{\text{NP},i}=1 \mid \boldsymbol{x}_i)\)
be propensity scores, which characterize the \(S_{\text{NP}}\) sample's
inclusion and participation mechanisms. Unlike \(\pi_{\text{P},i}\), the
\(\pi_{\text{NP},i}\) and \(d_{\text{NP},i}=1/\pi_{\text{NP},i}\) are
unknown. The description of the data is presented in a~more concise form
in Table \ref{tab-two-sources}.

\begin{table}[ht!]
    \centering
    \resizebox{\linewidth}{!}{
    \begin{tabular}{llcccc} 
    \hline
    Sample & ID & Inclusion ($I_{\text{NP}}$) & Design weight ($d$) & Covariates ($\boldsymbol{x}$) & Study variable ($y$) \\
    \hline
    Non-probability  & 1 & 1 & ? & $\checkmark$ & $\checkmark$ \\ 
    $S_{\text{NP}}$ & $\vdots$ & $\vdots$ & $\vdots$ & $\vdots$ & $\vdots$ \\
    & $n_{\text{NP}}$ & 1 & ? & $\checkmark$ & $\checkmark$ \\
    Probability  & $n_{\text{NP}}+1$ & 0 & $\checkmark$ & $\checkmark$ & ? \\
    $S_{\text{P}}$ & $\vdots$ & $\vdots$ & $\vdots$ & $\vdots$ & $\vdots$ \\ 
    & $n_{\text{NP}}+n_{\text{P}}$ & 0 & $\checkmark$ & $\checkmark$ & ? \\                                     
    \hline     
    \end{tabular}
    }
    \caption{Two-sample setting with no overlap between samples.}
    \label{tab-two-sources}
\end{table}

The goal is to estimate the finite population mean
\(\mu_{y} = N^{-1}\sum_{i=1}^{N} y_{i}\) of the target variable \(y\).
As values of \(y_{i}\) are not observed in the probability sample, they
cannot be used to estimate the target quantity. Instead, one could try
combining the non-probability and probability samples (or known
population totals) to estimate \(\mu_{y}\). Given the absence of a
universally accepted method for achieving this objective, assumptions
vary considerably, as outlined by \citet{wu2022statistical}. However,
the following are the main assumptions that apply to the majority of
methods presented in this section:

\begin{itemize}
\item[A1] $I_{\text{NP},i}$ and the study variable $y_{i}$ are independent given the set of covariates $\boldsymbol{x}_{i}$ (the missing at random mechanism).
\item[A2] All the units in the target population have non-zero PS, i.e., $\pi_{\text{NP},i}>0$, $i=1,2, \ldots, N$ (i.e., no coverage error).
\item[A3] The indicator variables $I_{\text{NP},1}, I_{\text{NP},2}, \ldots, I_{\text{NP},N}$ are independent given the set of auxiliary variables $\left(\boldsymbol{x}_1, \boldsymbol{x}_2, \ldots, \boldsymbol{x}_N\right)$ (i.e., no clustering).
\end{itemize}

Currently, we ignore overlap between \(S_{\text{NP}}\) and
\(S_{\text{P}}\), and assume no measurement error in \(y_i\) and the
fact that values of \(\boldsymbol{x}_i\) are known. The setting
presented in Table \ref{tab-two-sources} can also be extended to
calibrated \(d_{\text{P},i}\) weights (i.e., \(d_{\text{P},i}\) adjusted
for under-coverage, non-contact or non-response;
cf.~\cite{sarndal2005estimation}) but this requires additional
developments in the theory about the consistency of the MI, IPW and DR
estimators. In the next sections we briefly present the methods
implemented in the package.

\subsection{Prediction-based approach}\label{sec-prediction}

\subsubsection{Prediction estimators}\label{prediction-estimators}

In the prediction approach the following semi-parametric model for the
finite population is assumed:

\begin{equation}
E_{\xi}\left(y_i \mid \boldsymbol{x}_i\right)=m\left(\boldsymbol{x}_i, \boldsymbol{\beta}\right), \text { and } \quad V_{\xi}\left(y_i \mid \boldsymbol{x}_i\right)=v\left(\boldsymbol{x}_i\right) \sigma^2, \quad i=1,2, \ldots, N,
\label{eq-semipar-model}
\end{equation}

where the mean function \(m(\cdot,\cdot)\) and the variance \(v(\cdot)\)
have known forms, \(\boldsymbol{\beta}\) and \(\sigma^2\) are model
parameters and \(y_i\) are also assumed to be conditionally independent
given the \(\boldsymbol{x}_i\). The model in Equation
\ref{eq-semipar-model} is assumed to hold for all units in the
non-probability sample \(S_{\text{NP}}\). The parameters of the model in
Equation~\ref{eq-semipar-model} can be estimated using the quasi maximum
likelihood estimation method, which includes linear and non-linear
models, such as GLM. Let \(\boldsymbol{\beta}_0\) and \(\sigma^2_0\) be
the true values of the model parameters \(\boldsymbol{\beta}\) and
\(\sigma^2\) under the adopted model and \(\hat{\boldsymbol{\beta}}\) be
the quasi maximum likelihood estimator of \(\boldsymbol{\beta}_0\). Let
\(m_i=m(\boldsymbol{x}_i, \boldsymbol{\beta}_0)\) and
\(\hat{m}_i=m(\boldsymbol{x}_i, \hat{\boldsymbol{\beta}})\) be
calculated for all units \(i=1,...,N\). Under this setting, as
\citet{wu2022statistical} notes, there are two commonly used prediction
estimators:

\begin{equation}
\hat{\mu}_{y,\text{PR1}}=\frac{1}{N} \sum_{i=1}^N \hat{m}_i \quad \text { and } \quad \hat{\mu}_{y,\text{PR2}}=\frac{1}{N}\left\{\sum_{i \in S_{\text{NP}}} y_i-\sum_{i \in S_{\text{NP}}} \hat{m}_i+\sum_{i=1}^N \hat{m}_i\right\}.
\label{eq-pred-two-estimators}
\end{equation}

Under linear models, where
\(m(\boldsymbol{x}_i, \boldsymbol{\beta})=\boldsymbol{x}_i^{\top}\boldsymbol{\beta}\),
the two estimators in Equation~\ref{eq-pred-two-estimators} reduce to:

\begin{equation}
\hat{\mu}_{y,\text{PR1}}=\mu_{\boldsymbol{x}}^{\top} \hat{\boldsymbol{\beta}} \quad \text { and } \quad \hat{\mu}_{y,\text{PR2}}=\frac{n_{\text{NP}}}{N}\left(\overline{y}_{\text{NP}}-\overline{\boldsymbol{x}}_{\text{NP}}^{\top} \hat{\boldsymbol{\beta}}\right)+\mu_{\boldsymbol{x}}^{\top} \hat{\boldsymbol{\beta}},
\label{eq-pred-two-estimators-simplified}
\end{equation}

where \(\mu_{\boldsymbol{x}} = N^{-1}\sum_{i=1}^N\boldsymbol{x}_i\) is
the vector of the population means of the \(\boldsymbol{x}\) variables
and
\(\overline{\boldsymbol{x}}_{\text{NP}}=n_{\text{NP}}^{-1}\sum_{i \in S_{\text{NP}}}\boldsymbol{x}_i\)
is the vector of the simple means of \(\boldsymbol{x}\) from the
non-probability \(S_{\text{NP}}\) sample. If the linear model contains
an intercept and \(\hat{\boldsymbol{\beta}}\) is the ordinary least
squares estimator, then
\(\hat{\mu}_{y,\text{PR1}}=\hat{\mu}_{y,\text{PR2}}\).

The estimator in Equation~\ref{eq-pred-two-estimators-simplified} is
appealing as it only requires the non-probability sample
\(S_{\text{NP}}\) and reference population means (or totals and
population size \(N\)). If the population means are unknown, they can be
replaced by estimates provided by the reference probability sample
\(S_{\text{P}}\), i.e., \(\sum_{i=1}^N \hat{m}_i\) is replaced with
\(\sum_{i \in S_{\text{P}}} d_{\text{P},i}\hat{m}_i\) for Equation
\ref{eq-pred-two-estimators} and \(\mu_{\boldsymbol{x}}\) is replaced by
\(\hat{\mu}_{\boldsymbol{x}}=\hat{N}_{\text{P}}^{-1}\sum_{i \in S_{\text{P}}}d_{\text{P},i}\boldsymbol{x}_i\)
for Equation~\ref{eq-pred-two-estimators-simplified} where
\(\hat{N}_{\text{P}}=\sum_{i \in S_{\text{P}}}d_{\text{P},i}\).

\subsubsection{Mass imputation
estimators}\label{mass-imputation-estimators}

Model-based prediction estimators of \(\mu\) can be treated as mass
imputation estimators, since the information on \(y_i\) is missing
entirely in the reference probability sample \(S_{\text{P}}\) (but
\(\boldsymbol{x}_i\) is available) and \(y_i\) can be imputed based on
the non-probability sample as values of
\(\{ (y_i, \boldsymbol{x}_i), i \in \mathcal{I}_{\text{NP}}\}\) are
known. The general form of the MI estimator is given by:

\begin{equation}
\hat{\mu}_{y,\text{MI}}=\frac{1}{\hat{N}_{\text{P}}} \sum_{i \in S_{\text{P}}} d_{\text{P},i} y_i^*, 
\end{equation}

where \(y_i^*\) is the imputed value of \(y_i\) and
\(\hat{N}_{\text{P}}\) is defined as previously. Under deterministic
regression imputation, the \(\hat{\mu}_{y,\text{MI}}\) estimator reduces
to the estimators in Equation~\ref{eq-pred-two-estimators}.

There are several ways of imputing \(y_i^*\) and in the \pkg{nonprobsvy}
package we have implemented the following MI estimators: the
semi-parametric approach based on GLM (MI-GLM), kernel smoothing
(MI-NPAR), nearest neighbour matching (MI-NN) and predictive mean
matching (MI-PMM).

The properties of the MI-GLM estimator, where \(y_i^*\) are imputed with
\(\hat{m}_i\) from the semi-parametric model, were studied by
\citet{kim_combining_2021}. In the \pkg{nonprobsvy} package, we account
for the following GLM families: \code{gaussian()}, \code{binomial()} and
\code{poisson()}. The MI-NPAR estimator was proposed by
\citet{chen_nonparametric_2022} who studied its finite population
properties. Currently, we implement this approach using local polynomial
regression using \code{stats::loess()} function. In the next releases we
will allow users to use the \pkg{np} \citep{nppkg} or \pkg{KernSmooth}
\citep{KernSmooth} packages.

The MI-NN estimator was initially proposed by \citet{rivers2007sampling}
under the name sample matching and theoretical properties of the MI-NN
estimator for large non-probability samples (big data, i.e., covering a
significant part of the target population) were studied by
\citet{yang2021integration}. The basic idea of NN matching is as
follows:

\begin{enumerate}
\item for each unit $i$ in the probability sample $S_{\text{P}}$ find a~donor $j$ (or multiple donors) in the $S_{\text{NP}}$ sample based on some distance between $\boldsymbol{x}_i$ and $\boldsymbol{x}_j$,
\item use the matched values $y_j$ from $S_{\text{NP}}$ to impute missing $y_i$ values in the probability sample $S_{\text{P}}$. 
\end{enumerate}

Imputed values of \(y_i^*\) depend on the number of selected \(k\)
neighbours: for \(k=1\), the closest unit is selected, and for \(k>1\),
one can calculate a~simple average over a~vector of selected \(y\)
values. A detailed description of the procedure is presented in
Algorithm \ref{algo-2} in the Appendix. The MI-NN estimator suffers from
the curse of dimensionality, i.e., asymptotic bias of the MI estimator
increases as the number of covariates \(\boldsymbol{x}\) increases with
a fixed \(k\) \citep{abadie2006large, yang_asymptotic_2020}. The single
PMM imputation approach has been proposed to overcome this issue
\citep{chlebicki2025}.

In the PMM approach, matching is done using predicted values of
\(\hat{m}_i=m\left(\boldsymbol{x}_i, \hat{\boldsymbol{\beta}}\right)\)
instead of \(\boldsymbol{x}_i\), thus the NN algorithm is modified as
follows:

\begin{enumerate}
\item fit the model $m\left(\boldsymbol{x}_i, \boldsymbol{\beta}\right)$ to non-probability $S_{\text{NP}}$ sample, 
\item assign predicted values $\hat{m}_i$ to all units in $S_{\text{NP}}$ and $S_{\text{P}}$, 
\item match all units from the $S_{\text{P}}$ sample to donor units from the $S_{\text{NP}}$ sample based on $\hat{m}$ values. 
\end{enumerate}

The MI-PMM estimator is the same as in the NN approach but differs from
the multiple imputation PMM implemented in the \pkg{mice} package
\citep{mice}. In \pkg{nonprobsvy} we have implemented a~single
imputation method (i.e., one dataset is created). \citet{chlebicki2025}
studied properties of two variants of the MI-PMM estimator for
non-probability samples: matching predicted to predicted
(\(\hat{m}-\hat{m}\) matching; denoted as MI-PMM-A) and matching
predicted to observed (\(\hat{m}-y\) matching; denoted as MI-PMM-B).
Details of the procedure can be found in Algorithm \ref{algo-3} and
\ref{algo-4} in the Appendix. \citet{chlebicki2025} also prove the
consistency of the MI-PMM-A estimator under model misspecification,
i.e., the assumed model may differ from the true one.

\subsubsection{Variance estimators for the prediction
approach}\label{variance-estimators-for-the-prediction-approach}

Variance of the MI estimators can be estimated analytically or using the
bootstrap approach. The analytical estimator of the variance of the
MI-GLM estimator proposed by \citet[p. 950]{kim_combining_2021} contains
two components: \(\hat{V}_1\) (based on the information from both
samples \(S_{\text{NP}}\) and \(S_{\text{P}}\)) and \(\hat{V}_2\) (based
exclusively on the probability sample \(S_{\text{P}}\)). For the MI-NN
estimator \citet{yang2021integration} proposed a~variance estimator for
large \(S_{\text{NP}}\) samples, which reduces to the part for the
probability sample \(S_{\text{P}}\) (i.e., the design-based variance
estimator of the mean, which can easily be obtained from the
\pkg{survey} package) and a~version for smaller samples can be found in
\citet{chlebicki2025}. With respect to the variance of MI-PMM
estimators, \citet{chlebicki2025} propose formulas which are the same as
those for the MI-NN estimators. The analytical variance estimator for
the MI-NPAR was proposed by \citet{chen_nonparametric_2022}.

In the bootstrap approach each bootstrap replication \(b=1,...,B\)
consists of the following steps.

\begin{enumerate}
\item Independently:
  \begin{itemize}
  \item Draw a~simple random sampling with replacement from the non-probability sample $S_{\text{NP}}$ (using \code{base::sample()} function).
  \item Draw a~sample according to the declared sampling design from the probability sample $S_{\text{P}}$ (e.g., one can use the \code{as.svrepdesign()} function from the \pkg{survey} package).
  \end{itemize}
\item Estimate $\mu_{y, \text{MI}}^b$ using an appropriate approach (e.g., MI-GLM, MI-NN or MI-PMM).
\end{enumerate}

After obtaining \(B\) bootstrap replicates, estimate variance using the
following equation:

\begin{equation}
\hat{V}_{\text{boot}} = \frac{1}{B-1}\sum_{b=1}^B\left(\hat{\mu}^b_{y, \text{MI}} - \hat{\mu}_{y, \text{MI}}\right)^2.
\label{eq-var-bootstrap}
\end{equation}

The above approaches are applied when unit-level data from the
probability sample \(S_{\text{P}}\) are available. If this is not the
case and only population means (or totals and population size) are
available, we can estimate the variance of the \(\mu_{y,\text{MI-GLM}}\)
estimator using the first component \(\hat{V}_1\) of the
\citet{kim_combining_2021} variance estimator (replaced by the
survey-based population quantities, if available). To estimate the
variance of the MI-NN and MI-PMM estimators we only allow the bootstrap
approach with known population means. Note that the current version of
the \pkg{nonprobsvy} does not support the use of replicated weights in
the probability sample \(S_{\text{P}}\) for any of the estimators
discussed in this paper.

\subsection{Inverse probability weighting}\label{sec-ipw}

Inverse probability weighting, another popular estimation approach,
involves estimating PS given by
\(\pi_{\text{NP},i}=P\left(i \in S_{\text{NP}} \right)\). As in the case
of the prediction-based approach, there are two variants of the IPW
estimator, given by:

\begin{equation}
  \hat{\mu}_{y,\text{IPW-HT}}=\frac{1}{N} \sum_{i \in S_{\text{NP}}} \frac{y_i}{\hat{\pi}_{\text{NP},i}} \quad \text{and} \quad 
  \hat{\mu}_{y,\text{IPW-H\'{a}jek}}=\frac{1}{\hat{N}_{\text{NP}}} \sum_{i \in S_{\text{NP}}} \frac{y_i}{\hat{\pi}_{\text{NP},i}},
\label{eq-ipw-estimators}
\end{equation}

where the \(\hat{\mu}_{y,\text{IPW-HT}}\) is an adjustment of the
Horvitz-Thompson estimator, and the
\(\hat{\mu}_{y,\text{IPW-H\'{a}jek}}\) is an adjustment of the Hájek
estimator, where the estimated population size is given by
\(\hat{N}_{\text{NP}} = \sum_{i \in S_{\text{NP}}} \hat{\pi}_{\text{NP},i}^{-1}\).
The use of this estimator with respect to non-probability samples is
discussed by \citet{lee2006propensity} and
\citet[Chapter 13]{biffignandi2021handbook} and there are several
approaches to using propensity scores along with alternative versions of
the weights \citep[cf.][Section 3]{elliott_inference_2017}. In an
article by \citet{chen2020doubly} dedicated to the properties of the
estimators in Equation~\ref{eq-ipw-estimators}, the authors proved their
consistency and derived their closed form versions.
\citet[Section 4.2]{wu2022statistical} argues that the
\(\hat{\mu}_{y,\text{IPW-H\'{a}jek}}\) estimator performs better than
\(\hat{\mu}_{y,\text{IPW-HT}}\) even if the population size is known.

The construction of the IPW estimator involves two steps:

\begin{enumerate}
\item estimating the PS, 
\item deriving $d_{\text{NP},i}$ weights (which, in our case, are equal to $\pi_{\text{NP},i}^{-1}$) 
\end{enumerate}

To estimate the propensity scores
\(\pi_{\text{NP},i}=\pi(\boldsymbol{x}_i, \boldsymbol{\gamma})\) one can
use the likelihood approach assuming that the information about
\(\boldsymbol{x}_i\) is available for each unit in the population given
by Equation~\ref{eq-ipw-loglik}.

\begin{equation}
\begin{aligned}
\ell(\boldsymbol{\gamma}) &= \log\left\{\prod_{i=1}^N \left(\pi_{\text{NP,i}}\right)^{I_{\text{NP},i}}\left(1-\pi_{\text{NP},i}\right)^{1-I_{\text{NP},i}}\right\} \\
&= \sum_{i \in S_{\text{NP}}} \log \left\{\frac{\pi\left(\boldsymbol{x}_i, \boldsymbol{\gamma}\right)}{1-\pi\left(\boldsymbol{x}_i, \boldsymbol{\gamma}\right)}\right\}+\sum_{i=1}^N \log \left\{1-\pi\left(\boldsymbol{x}_i, \boldsymbol{\gamma}\right)\right\}.
\end{aligned}
\label{eq-ipw-loglik}
\end{equation}

In practice, a~function of this form cannot be used because not all
units from the population are observed. A more realistic approach
consists in using a~reference probability sample \(S_{\text{P}}\), which
means that the second component of the equation in Equation
\ref{eq-ipw-loglik} is replaced, yielding a~pseudo log-likelihood
function given by Equation~\ref{eq-ipw-pseudo-loglik}

\begin{equation}
\ell^*(\boldsymbol{\gamma}) = \sum_{i \in S_{\text{NP}}} \log \left\{\frac{\pi\left(\boldsymbol{x}_i, \boldsymbol{\gamma}\right)}{1-\pi\left(\boldsymbol{x}_i, \boldsymbol{\gamma}\right)}\right\}+ \sum_{i \in S_{\text{P}}} d_{\text{P},i} \log \left\{1-\pi\left(\boldsymbol{x}_i, \boldsymbol{\gamma}\right)\right\}.
\label{eq-ipw-pseudo-loglik}
\end{equation}

The maximum pseudo-likelihood estimator \(\hat{\boldsymbol{\gamma}}\)
can be obtained as the solution to the pseudo score equation, which,
under the logit function assumed for \(\pi_{\text{NP},i}\), is given by
Equation~\ref{eq-ipw-solution-mle}

\begin{equation}
\boldsymbol{U}(\boldsymbol{\gamma}) = \sum_{i \in S_{\text{NP}}} \boldsymbol{x}_i - \sum_{i \in S_{\text{P}}} d_{\text{P},i} \pi(\boldsymbol{x}_i, \boldsymbol{\gamma}) \boldsymbol{x}_i.
\label{eq-ipw-solution-mle}
\end{equation}

In general, pseudo score functions
\(\boldsymbol{U}(\boldsymbol{\gamma})\) for true values of model
parameters \(\boldsymbol{\gamma}_0\) are unbiased under the joint
\(q p\) randomization in the sense that
\(E_{q p}\left\{\boldsymbol{U}\left(\boldsymbol{\gamma}_0\right)\right\}=\boldsymbol{0}\),
which implies that the estimator \(\hat{\boldsymbol{\gamma}}\) is
consistent for \(\boldsymbol{\gamma}_0\) \citep{wu2022statistical}.

The terms in Equation~\ref{eq-ipw-solution-mle} can be replaced by
general estimation equations. Let
\(\boldsymbol{h}(\boldsymbol{x}, \boldsymbol{\gamma})\) be a
user-specified vector of functions with the same dimension of
\(\boldsymbol{\gamma}\) and \(\boldsymbol{G}(\boldsymbol{\gamma})\) is
defined as

\begin{equation}
\label{gee}
\boldsymbol{G}(\boldsymbol{\gamma})=\sum_{i \in S_{\text{NP}}} \boldsymbol{h}\left(\boldsymbol{x}_i, \boldsymbol{\gamma}\right)-\sum_{i \in S_{\text{P}}} d_{\text{P},i} \pi\left(\boldsymbol{x}_i, \boldsymbol{\gamma}\right) \boldsymbol{h}\left(\boldsymbol{x}_i, \boldsymbol{\gamma}\right),
\end{equation}

then solving for \(\boldsymbol{G}(\boldsymbol{\gamma})=\boldsymbol{0}\)
with the chosen parametric form of \(\pi_{\text{NP},i}\) and the chosen
\(\boldsymbol{h}(\boldsymbol{x},\boldsymbol{\gamma})\) produces the
consistent estimator of \(\hat{\boldsymbol{\gamma}}\). In the
literature, the most commonly considered functions are
\(\boldsymbol{h}\left(\boldsymbol{x}_i, \boldsymbol{\gamma}\right) = \boldsymbol{x}_i\)
and
\(\boldsymbol{h}\left(\boldsymbol{x}_i, \boldsymbol{\gamma}\right) = \boldsymbol{x}_i \pi\left(\boldsymbol{x}_i, \boldsymbol{\gamma}\right)^{-1}\).
Note that if the function
\(\boldsymbol{h}\left(\boldsymbol{x}_i, \boldsymbol{\gamma}\right)=\boldsymbol{x}_i\),
then \(\boldsymbol{G}\) reduces to \(\boldsymbol{U}\) and for the second
variant of the \(\boldsymbol{h}\) function we get

\begin{equation}
\boldsymbol{G}(\boldsymbol{\gamma}) = \sum_{i \in S_{\text{NP}}} \frac{\boldsymbol{x}_i}{\pi\left(\boldsymbol{x}_i, \boldsymbol{\gamma}\right) }-\sum_{i \in S_{\text{P}}} d_{\text{P},i} \boldsymbol{x}_{i},
\label{eq-ipw-solution-gee}
\end{equation}

which can be viewed as calibrated IPW. Equation
\ref{eq-ipw-solution-gee} only requires the knowledge of population
totals for auxiliary variables \(\boldsymbol{x}\). Moreover, the use of
Equation~\ref{eq-ipw-solution-gee} yields a~doubly robust estimator
under the assumption that the outcome model is linear
\citep[cf.][]{kim_theory_2012}.

\subsubsection{Variance estimators for the inverse probability weighting
approach}\label{variance-estimators-for-the-inverse-probability-weighting-approach}

\citet[Section 3.2]{chen2020doubly} derived asymptotic variance
estimators for both IPW estimators in Equation~\ref{eq-ipw-estimators}
and presented the plug-in variance estimator for the
\(\hat{\mu}_{y,\text{IPW-H\'{a}jek}}\) estimator assuming logistic
regression. In the package we have implemented this approach for
\code{"logit"}, \code{"probit"} and \code{"cloglog"} link functions of
the \code{binomial()} family. We refer the reader to
\citet[Section 6.2]{wu2022statistical} and
\citet[Chapter 3]{chrostowski2024statistical} for more details on how
these estimators are derived based on the general estimating equations
approach.

Another approach is to use a~bootstrap, which is essentially the same as
the one presented in Equation~\ref{eq-var-bootstrap}, where
\(\hat{\mu}_{y}\) is replaced by one of the estimators in Equation
\ref{eq-ipw-estimators}.

\subsection{Doubly robust approach}\label{sec-dr-approach}

The IPW and MI estimators are suited to correctly specified models for
PS and outcome regression models, respectively. The DR approach was
proposed to improve robustness and efficiency
\citep[cf.][]{robins1994estimation}. It incorporates a~prediction model
for the response variable \(y_i\) and a~PS model for participation
\(I_{\text{NP},i}\). This approach is doubly robust in the sense that
the DR estimator remains consistent even if one of the models is
misspecified. We need to consider a~joint randomization approach
involving a~non-probability sample \(S_{\text{NP}}\) and a~probability
sample \(S_{\text{P}}\) and DR inference is conducted within the \(qp\)
or \(\xi p\) framework without specifying which one is correct. The
general formula for the DR estimator is given by

\[
\tilde{\mu}_{y,\text{DR}} = \frac{1}{N}\sum_{i \in S_{\text{NP}}}\frac{y_i - m_i}{\pi_{\text{NP},i}} + \frac{1}{N}\sum_{i =1}^N m_i,
\]

where \(\pi_{\text{NP},i}\) and \(m_i\) are defined as previously. In
the next sections we discuss two approaches to the DR estimation.

\subsubsection{Parameters estimated
separately}\label{parameters-estimated-separately}

\citet{chen2020doubly} proposed two DR estimators given in Equation
\ref{eq-dr-known-n} and Equation~\ref{eq-dr-estimated-n} assuming that
the population size is either known or estimated:

\begin{equation}
\hat{\mu}_{y,\text{DR-HT}}=\frac{1}{N} \sum_{i \in S_{\text{NP}}} d_{\text{NP},i}\left\{y_i-m\left(\boldsymbol{x}_i, \hat{\boldsymbol{\beta}}\right)\right\}+\frac{1}{N} \sum_{i \in S_{\text{P}}} d_{\text{P},i} m\left(\boldsymbol{x}_i, \hat{\boldsymbol{\beta}}\right),
\label{eq-dr-known-n}
\end{equation}

and

\begin{equation}
\hat{\mu}_{y,\text{DR-H\'{a}jek}}=\frac{1}{\hat{N}_{\text{NP}}} \sum_{i \in S_{\text{NP}}} d_{\text{NP},i}\left\{y_i-m\left(\boldsymbol{x}_i, \hat{\boldsymbol{\beta}}\right)\right\}+\frac{1}{\hat{N}_{\text{P}}} \sum_{i \in S_{\text{P}}} d_{\text{P},i} m\left(\boldsymbol{x}_i, \hat{\boldsymbol{\beta}}\right),
\label{eq-dr-estimated-n}
\end{equation}

where
\(d_{\text{NP},i}=\pi\left(\boldsymbol{x}_i, \hat{\boldsymbol{\gamma}}\right)^{-1}\),
\(\hat{N}_{\text{NP}}=\sum_{i \in S_{\text{NP}}} d_{\text{NP},i}\) and
\(\hat{N}_{\text{P}}=\sum_{i \in S_{\text{P}}} d_{\text{P},i}\). The
estimator in Equation~\ref{eq-dr-estimated-n}, including the estimated
population size, has a~better performance in terms of bias and the mean
squared error and should be used in practice. However, the main
limitation is associated with variance estimation, which is discussed at
the end of this section.

\citet{chen2020doubly} suggested constructing the estimators in Equation
\ref{eq-dr-known-n} or \ref{eq-dr-estimated-n} based on the two models
estimated separately, which is an attractive option, since one can
specify a~different number of variables for the propensity and outcome
model. An alternative approach, proposed by \citet{yang_doubly_2020},
and similar to the one described by \citet{kim2014doubly}, is discussed
in the next section.

\subsubsection{Minimization of the bias for doubly robust
methods}\label{minimization-of-the-bias-for-doubly-robust-methods}

\citet{yang_doubly_2020} discussed variable selection for a
high-dimensional setting and noted that the asymptotic bias of the
estimator, which can increase, cannot be controlled. Therefore,
according to \citet{yang_doubly_2020}, the idea is to develop equations
that can be used to estimate the \(\boldsymbol{\beta}\) and
\(\boldsymbol{\gamma}\) parameters based on the bias of the population
mean estimator. In this way the parameters can be estimated in a~single
step, rather than in two separate steps. First, the authors derived the
asymptotic bias of the \(\hat{\mu}_{\text{DR}}\), assuming
\(\boldsymbol{h}(\boldsymbol{x}_i, \boldsymbol{\gamma})=\boldsymbol{x}_i\pi(\boldsymbol{x}_i, \boldsymbol{\gamma})^{-1}\)
for the IPW estimator, which is given by Equation~\ref{eq-dr-bias}

\begin{equation}
\begin{aligned}
\operatorname{bias}\left(\hat{\mu}_{\text{DR}}\right) = E|\hat{\mu}_{\text{DR}}-\mu| &=
E\left\{\frac{1}{N} \sum_{i=1}^N\left\{\frac{I_{\text{NP},i}}{\pi\left(\boldsymbol{x}_i, \boldsymbol{\gamma}\right)} - 1\right\}\left\{y_i-m\left(\boldsymbol{x}_i, \boldsymbol{\beta}\right)\right\} \right\}\\
& + E\left\{\frac{1}{N} \sum_{i=1}^N\left(I_{\text{P},i} d_{\text{P},i}-1\right) m\left(\boldsymbol{x}_i, \boldsymbol{\beta}\right)\right\}.
\end{aligned}
\label{eq-dr-bias}
\end{equation}

The goal of this approach is to minimize
\(\operatorname{bias}\left(\hat{\mu}_{\text{DR}}\right)^2\), which
consists in solving the following system of empirical equations:

\begin{equation}
\left(\begin{array}{c}
\sum_{i=1}^N I_{\text{NP},i}\left\{\frac{1}{\pi\left(\boldsymbol{x}_i, \boldsymbol{\gamma}\right)}-1\right\}\left\{y_i-m\left(\boldsymbol{x}_i, \boldsymbol{\beta}\right)\right\} \boldsymbol{x}_i \\
\sum_{i=1}^N \frac{I_{\text{NP},i}}{\pi\left(\boldsymbol{x}_i, \boldsymbol{\gamma}\right)} \dot{m}\left(\boldsymbol{x}_i, \boldsymbol{\beta}\right) -\sum_{i \in S_{\text{P}}} d_{\text{P},i} \dot{m}\left(\boldsymbol{x}_i, \boldsymbol{\beta}\right)
\end{array}\right) = \boldsymbol{0},
\label{eq-bias-min}
\end{equation}

where
\(\dot{m}\left(\boldsymbol{x}_i, \boldsymbol{\beta}\right)=\frac{\partial m\left(\boldsymbol{x}_i, \boldsymbol{\beta}\right)}{\partial \boldsymbol{\beta}}\).
The system in Equation~\ref{eq-bias-min} can be solved using the
Newton--Raphson method. This approach, without variable selection, is
equivalent to that proposed by \citet{kim2014doubly} and was extensively
discussed by \citet{chen2020doubly} and \citet{wu2022statistical} in the
context of estimating parameters and the variance of the DR estimator.
The main limitation of this approach is the possibility that a~solution
to Equation~\ref{eq-bias-min} may not exist unless the two sets of
covariates used in the outcome regression model and the PS model have
the same dimensions. That is why \citet{yang_doubly_2020} suggested
using this approach on the union of variables from both models (e.g.,
after variable selection).

In the \pkg{nonprobsvy} package we have implemented these approaches not
only for
\(\boldsymbol{h}(\boldsymbol{x}_i, \boldsymbol{\gamma})=\boldsymbol{x}_i \pi\left(\boldsymbol{x}_i, \boldsymbol{\gamma}\right)^{-1}\)
but also for
\(\boldsymbol{h}(\boldsymbol{x}_i, \boldsymbol{\gamma})=\boldsymbol{x}_i\)
and various link functions for the propensity score model. As noted in
the beginning, the choice of either Equation~\ref{eq-dr-known-n} or
\ref{eq-dr-estimated-n} results in a~different approach to estimating
variance, which is discussed in the next section.

\subsubsection{Variance estimators for the doubly robust
approach}\label{variance-estimators-for-the-doubly-robust-approach}

\citet{yang_doubly_2020} derived a~closed form estimator for Equation
\ref{eq-dr-known-n} but this requires the knowledge of the population
and bias correction to obtain a~valid estimator for
\(V_{\xi p}\left(\hat{\mu}_{y,\text{DR}}-\mu_y\right)\) under the
outcome regression model \(\xi\). A doubly robust variance estimator for
\(\hat{\mu}_{y,\text{DR-H\'{a}jek}}\) given by Equation
\ref{eq-dr-estimated-n} is not yet available in the literature. In the
package, to offer the analytical variance estimator of
\(\hat{\mu}_{y,\text{DR-H\'{a}jek}}\) we simply replace \(N\) with
estimated \(\hat{N}_{\text{NP}}\) and \(\hat{N}_{\text{P}}\) but we urge
caution when using this approach.

Alternatively, one can use the bootstrap approach.
\citet{chen2020doubly} demonstrated that the bootstrap approach
presented in Section \ref{sec-prediction} performs well in terms of the
coverage rate when one of the working models is correctly specified.
This is why this approach is recommended for all users.

\subsection{Variable selection algorithms}\label{sec-varsel}

\citet{yang_asymptotic_2020} point out that it is crucial to use
variable selection techniques during estimation, especially when dealing
with high-dimensional non-probability samples. Variable selection not
only improves model stability and computational feasibility, but also
reduces variance, which can increase when irrelevant auxiliary variables
are included.

The most popular approaches described in the literature are penalization
methods, such as least absolute shrinkage and selection operator
(LASSO), smoothly clipped absolute deviation (SCAD) or minimax concave
penalty (MCP), which, thanks to the appropriate loss functions, shrink
the coefficients in variables that have no significant effect on the
dependent variable \citep[cf.][]{tibshirani1996regression, ncvreg}.

The selection procedure for non-probability methods works in a~similar
way, with loss functions modified to account for external data sources,
such as sample or population totals or averages. In particular, the
technique consists of two steps:

\begin{enumerate}
\item select the relevant variables using an appropriately constructed loss function (and possibly using the approach shown in Equation~\ref{eq-bias-min} to obtain the final estimates of the model parameters),
\item construct the selected estimator using variables selected from the first step. 
\end{enumerate}

For instance, \citet{yang_doubly_2020} used Equation
\ref{eq-varsel-loss-mi} as a~loss function for estimating outcome
equation parameters:

\begin{equation}  
\operatorname{Loss}\left(\lambda_{\boldsymbol{\beta}}\right)=\sum_{i=1}^N I_{\text{NP},i}\left[y_i-m\left\{\boldsymbol{x}_i, \boldsymbol{\beta}(\lambda_{\boldsymbol{\beta}})\right\}\right]^2,
\label{eq-varsel-loss-mi}
\end{equation}

where
\(m\left\{\boldsymbol{x}_i, \boldsymbol{\beta}(\lambda_{\boldsymbol{\beta}})\right\}\)
is the penalized function for the \(\boldsymbol{\beta}\) parameters with
a tuning parameter \(\lambda_{\boldsymbol{\beta}}\) and loss functions
for the PS function presented in Table \ref{tab-varsel-loss-ipw}, where
\(\lambda_{\boldsymbol{\gamma}}\) is the tuning parameter and
\(\boldsymbol{x}_{i,l}\) denotes \(l\)-th component of
\(\boldsymbol{x}_{i}\).

\begin{table}[ht!]
\centering
\begin{tabular}{ll}
\hline
$\boldsymbol{h}(\boldsymbol{x}_i, \boldsymbol{\gamma})$ function &  Loss function\\
\hline
$\boldsymbol{x}_i$ & $\sum_{l=1}^L\left(\sum_{i=1}^N\left[I_{\text{NP},i} - \frac{I_{\text{P},i}\pi\left\{\boldsymbol{x}_i, \boldsymbol{\gamma}(\lambda_{\boldsymbol{\gamma}})\right\}}{\pi_{\text{P},i}}\right] \boldsymbol{x}_{i,l}\right)^2$\\
\hline
$\boldsymbol{x}_i \pi(\boldsymbol{x}_i, \boldsymbol{\gamma})^{-1}$ & $\sum_{l=1}^L\left(\sum_{i=1}^N\left[\frac{I_{\text{NP},i}}{\pi\left\{\boldsymbol{x}_i, \boldsymbol{\gamma}(\lambda_{\boldsymbol{\gamma}})\right\}}-\frac{I_{\text{P},i}}{\pi_{\text{P},i}}\right] \boldsymbol{x}_{i,l}\right)^2$ \\
\hline
\end{tabular}
\caption{Loss functions for the PS function depending on the $\boldsymbol{h}(\cdot,\cdot)$ function.}
\label{tab-varsel-loss-ipw}
\end{table}

\citet{yang_doubly_2020} only discussed the SCAD penalty and the
\(\boldsymbol{h}(\boldsymbol{x}_i, \boldsymbol{\gamma})=\boldsymbol{x}_i\)
function for the DR estimator. In the \pkg{nonprobsvy} package we have
extended this approach to the first variant of
\(\boldsymbol{h}(\boldsymbol{x}_i, \boldsymbol{\gamma})\), shown in the
first row of Table \ref{tab-varsel-loss-ipw}, and have allowed the user
to select other link functions for the \(\pi_{\text{NP},i}\),
implemented other penalty functions and extended the possibility of
selecting variables to MI and IPW estimators. In the next section we
discuss how to define the approaches presented above.

\section{The main function and the package
functionalities}\label{sec-package}

\subsection[The nonprob function]{The \code{nonprob} function}

The \pkg{nonprobsvy} package is built around the \code{nonprob()}
function. The main design objective was to make the use of
\code{nonprob} as similar as possible to standard R functions for
fitting statistical models, such as \code{stats::glm()}, while
incorporating survey design features from the \pkg{survey} package. The
most important arguments are given in Table \ref{tab-arguments-nonprob}
and the obligatory ones include \code{data} as well as one of the
following three -- \code{selection}, \code{outcome}, or \code{target} --
depending on which method has been selected. In the case of
\code{outcome} and \code{target} multiple \(y\) variables can be
specified.

\begin{table}[ht!]
\centering
\scriptsize
\resizebox{\linewidth}{!}{
\begin{tabular}{p{3.25cm}p{10.25cm}}
\hline
Argument & Description \\
\hline
   \code{data} & a~\code{data.frame} with data from the non-probability sample of size $n_{\text{NP}}$. The data frame must contain all variables referenced in the \code{selection}, \code{outcome}, and \code{target} formulas\\
   \code{selection} & a~\code{formula} for the selection equation (e.g., \code{\string~x1 + x2}). This formula defines the PS model (for the IPW estimator)\\
   \code{outcome} & a~\code{formula} for the outcome equation (e.g., \code{y1 + y2 \string~ x1 + x2}). We allow for multiple target variables under the same prediction model (for the MI and DR estimator)\\
   \code{target} & a~\code{formula} with target variables (e.g., \code{\string~y1 + y2 + y3})\\
   \code{svydesign} & an optional \code{svydesign2} object containing the probability sample of size $n_{\text{P}}$. Must include all variables specified in \code{selection}, \code{outcome}, and \code{target} formulas. Used for calibration and validation of the non-probability sample estimates\\
   \code{pop\_totals}, 
   \code{pop\_means}, 
   \code{pop\_size} & an optional named \code{vector} with population totals or means of the covariates and population size. For \code{pop\_totals} and \code{pop\_means}, length must equal the number of covariates in the selection model, with names matching variable names in the \code{data} frame\\
   \code{method\_selection} & a~link function for the PS (\code{c("logit", "probit", cloglog")})\\
   \code{method\_outcome} & specification of the MI approach (\code{c("glm", "nn", "pmm", "npar")}) \\
   \code{family\_outcome} & a~GLM family for the MI approach (\code{c("gaussian", "binomial", "poisson")}) \\
   \code{subset} & an optional \code{vector} specifying a~subset of observations to be used in the fitting process\\
   \code{strata} & an optional parameter for estimation for sub-populations (currently not supported) \\
   \code{case\_weights} & an optional \code{vector} of prior case (frequency) weights to be used in the fitting process\\
   \code{na\_action} & a~\code{function} indicating what to do with \code{NA}'s (default \code{na.omit}) \\
   \code{control\_selection},
   \code{control\_outcome}, 
   \code{control\_inference} & control functions with parameters for PS, the outcome model and variance estimation, respectively \\
   \code{start\_selection}, \code{start\_outcome}  & an optional \code{vector} with starting values for the parameters of the PS and the outcome equation\\
   \code{verbose} & a~\code{logical} value indicating if information should be printed\\
   \code{se} & a~\code{logical} value indicating whether to calculate and return the standard error of the estimated mean\\
   \code{...} & additional optional argument for further development (currently not supported) \\
\hline
\end{tabular}
}
\caption{A description of the \code{nonprob()} function arguments.}
\label{tab-arguments-nonprob}
\end{table}

The package allows the user to provide either reference population data
(via the \code{pop\_totals}, or \code{pop\_means} and \code{pop\_size})
or a~probability sample declared by the \code{svydesign} argument
(\code{svydesign2} class from the \pkg{survey} package). The
\code{nonprob()} function is used to specify inference methods through
the \code{selection} and \code{outcome} arguments.

If a~\code{svydesign2} object is provided and the \code{selection}
argument is specified, then the IPW estimators are used (by default
parameters of the PS model employ Equation~\ref{eq-ipw-solution-mle}),
if the \code{outcome} argument is specified, then the MI approach is
used (the default option is the MI-GLM with the \code{gaussian} family)
and if both are specified, then the DR approach is applied (parameters
\((\boldsymbol{\beta}, \boldsymbol{\gamma})\) are estimated separately
and the Equation~\ref{eq-dr-estimated-n} is used). A particular
inference method is selected through the \code{method\_selection},
\code{method\_outcome}, \code{family\_outcome},
\code{control\_selection} and \code{control\_outcome} arguments. The
variance estimation method is selected through the
\code{control\_inference} argument.

Resulting object of class \code{nonprob} is a~list that contains the
following (most important) elements:

\begin{itemize}
\item \code{data} -- a~\code{data.frame} containing the non-probability sample.
\item \code{X} -- a~\code{matrix} containing both samples.
\item \code{y} -- a~\code{list} containing all variables declared in either the \code{target} or \code{outcome} arguments.
\item \code{R} -- a~numeric \code{vector} informing about inclusion in the non-probability sample.
\item \code{ps_scores} -- propensity scores or \code{NULL} (for the MI estimators).
\item \code{ipw_weights} -- inverse probability weights or \code{NULL} (for the MI estimators).
\item \code{output} - a~\code{data.frame} containing point and standard error estimates.
\item \code{confidence_interval} - a~\code{data.frame} containing confidence intervals for the mean.
\item \code{outcome} -- a~\code{list} of results for each \code{outcome} model.
\item \code{selection} -- a~\code{list} of results for the \code{selection} model.
\item \code{svydesign} -- a~\code{svydesign2} object passed by the \code{svydesign} argument.
\item \code{ys_rand_pred, ys_nons_pred} -- a~\code{list} of predicted values from the MI model.
\end{itemize}

\subsection{Controlling the type of
estimators}\label{controlling-the-type-of-estimators}

The \code{control_out()} function can be used to specify various aspects
of the estimation process, including the variable selection methods
(through different \code{penalty} options like SCAD, LASSO, and MCP with
their respective tuning parameters defined in the same way as in the
\code{control_sel()} function; cf.~\citet{yang_doubly_2020}), and
detailed configuration for NN, PMM and NPAR approaches (using parameters
like \code{pmm_match_type}, \code{pmm_weights}, \code{pmm_k_choice} or
\code{npar_loess}). For NN and PMM approaches we use the \pkg{RANN}
package \citep{rann-pkg} with the kd-tree algorithm and the Euclidean
distance as default. We currently do not support other distances (e.g.,
Gower). Table \ref{tab-control-out-examples} presents example usage of
the \code{control_out()} function for three types of MI estimators.

\begin{table}[ht!]
\centering
\small
\begin{tabular}{p{4cm}p{11cm}}
\hline
Estimator & Declaration with \code{control_out()} \\
\hline
MI-GLM with the LASSO penalty and 5 folds & 
\code{nonprob(outcome = y1 ~ x1 + x2, \newline
data = df, svydesign = prob, \newline
control_outcome = control_out(penalty = "lasso", \newline nfolds = 5))}\\
MI-NN with the \code{bd} algorithm & 
\code{nonprob(outcome = y1 ~ x1 + x2, \newline
data = df, svydesign = prob, \newline
control_outcome = control_out(treetype = "bd"))}\\
MI-PMM-B with $k=3$ & 
\code{nonprob(outcome = y1 ~ x1 + x2,  \newline
data = df, svydesign = prob, \newline
control_outcome = control_out(k = 3, pmm_match_type = 2))}\\
\hline
\end{tabular}
\caption{Example declarations of the MI estimators.}
\label{tab-control-out-examples}
\end{table}

The \code{control_sel()} function provides essential control parameters
for fitting the selection model in the \code{nonprob()} function. It
allows users to select between MLE given by Equation
\ref{eq-ipw-solution-mle} or GEE defined in Equation
\ref{eq-ipw-solution-gee} through the \code{est_method} argument,
specify the \(\boldsymbol{h}(\cdot, \cdot)\) function through the
\code{gee_h_fun} argument, specify the optimizer (\code{optimizer}
argument) and which variable selection method should be applied (using
different penalty functions like SCAD, lasso, and MCP by specifying the
\code{penalty} argument) along with parameters (e.g., the number of
folds through the \code{nfolds} argument). The parameters of the PS for
the calibrated IPW is estimated by using the \pkg{nleqslv} package and
fitting parameters (arguments starting with the \code{nleqslv_*}). Table
\ref{tab-control-sel-examples} presents example usage of the
\code{control_sel()} function for two types of IPW estimators.

\begin{table}[ht!]
\centering
\small
\begin{tabular}{p{4cm}p{11cm}}
\hline
Estimator & Declaration with \code{control_sel()} \\
\hline
Calibrated IPW & 
\code{nonprob(selection = ~ x1 + x2, target = ~y1, \newline
data = df, svydesign = prob,\newline
control_selection = control_sel(est_method = "gee"))}\\
IPW with the MCP penalty and 5 folds & 
\code{nonprob(selection = ~ x1 + x2, target = ~y1, \newline
data = df, svydesign = prob, \newline
control_selection = control_sel(penalty = "MCP", \newline nfolds = 5))}\\
\hline
\end{tabular}
\caption{Example declarations of the IPW estimators.}
\label{tab-control-sel-examples}
\end{table}

\subsection{Controlling variance
estimation}\label{controlling-variance-estimation}

Finally, the \code{control_inf()} function configures the parameters for
variance estimation in the \code{nonprob()} function. It allows users to
specify whether the analytical or bootstrap approach should be used (the
\code{var_method} argument), whether the variable selection method
should be applied (the \code{vars_selection} argument) and what type of
bootstrap should be applied for the probability sample (the
\code{rep_type} argument). This function is also used to specify the
inference procedure for the DR approach: if a~union of variables should
be applied (the \code{vars_combine} argument) and if the bias correction
(the \code{bias_correction} argument) should be applied after variable
selection (the \code{vars_selection}) and variable combination. Table
\ref{tab-control-inf-examples} presents example usage of the
\code{control_inf()} function for the IPW and DR estimators.

\begin{table}[ht!]
\centering
\small
\begin{tabular}{p{4cm}p{11cm}}
\hline
Estimator & Declaration with the \code{control_sel()} \\
\hline
Calibrated IPW with variable selection, bootstrap and $B=50$ & 
\code{nonprob(selection = ~ x1 + x2, target = ~y1, \newline
data = df, svydesign = prob, \newline
control_selection = control_sel(est_method = "gee"), \newline
control_inference = control_inf(vars_selection = TRUE,  \newline
var_method = "bootstrap", rep_type = "subbootstrap", \newline B = 50))}\\
The DR with the SCAD penalty, 5 folds and bias correction & 
\code{nonprob(selection = ~ x1 + x2, outcome = y1 ~ x1 + x2,  \newline
data = df, svydesign = prob, \newline
control_selection = control_sel(penalty = "SCAD", \newline nfolds = 5), \newline
control_inference = control_inf(vars_selection = TRUE, bias_correction = TRUE,  
vars_combine = TRUE))}\\
\hline
\end{tabular}
\caption{Example declarations of the IPW estimators.}
\label{tab-control-inf-examples}
\end{table}

In the next sections we present a~case study illustrating the process of
integrating a~non-probability sample with a~reference probability
sample. We present various estimators and compare them. Finally, we
describe more advanced options available in the package.

\section{Data analysis example}\label{sec-data-analysis}

\subsection{Description of the data}\label{description-of-the-data}

The package can be installed in the standard manner using:

\begin{CodeChunk}
\begin{CodeInput}
R> install.packages("nonprobsvy")
\end{CodeInput}
\end{CodeChunk}

Before we explain the case study let us first load the necessary
packages for this paper.

\begin{CodeChunk}
\begin{CodeInput}
R> library("nonprobsvy")
R> library("ggplot2")  
\end{CodeInput}
\end{CodeChunk}

The goal of the case study was to integrate survey (\code{jvs}) and
administrative (\code{admin}) data about job vacancies in Poland. The
first source, the job vacancy survey (JVS), contains 6,523 units. The
JVS provides a~probability sample drawn according to a~stratified
sampling design. More details can be found in \cite{jvs2022}. The
dataset contains information about the industry code (14 levels, the
\code{nace} column), \code{region} (16 levels), sector (2 levels, the
\code{private} column), company size (3 levels: Small up to 9, Medium
10-49 and Large 50+) and the final weight (i.e., the design weight
corrected for non-contact and non-response), which is treated as the
\(d\) weight.

\begin{CodeChunk}
\begin{CodeInput}
R> data("jvs")
R> head(jvs)
\end{CodeInput}
\begin{CodeOutput}
   id private size nace region weight
1 j_1       0    L    O     14      1
2 j_2       0    L    O     24      6
3 j_3       0    L  R.S     14      1
4 j_4       0    L  R.S     14      1
5 j_5       0    L  R.S     22      1
6 j_6       0    M  R.S     26      1
\end{CodeOutput}
\end{CodeChunk}

Since the \pkg{nonprobsvy} package relies on the functionalities of the
\pkg{survey} package, we need to define the \code{svydesign2} object via
the \code{svydesign} function, as shown below. The dataset does not
contain the true stratification variable, so we use a~simplified version
by specifying \texttt{\textasciitilde{}\ size\ +\ nace\ +\ region};
similarly, since we do not have information regarding non-response and
its correction, we simply assume that the \texttt{weight} sums up to the
population size.

\begin{CodeChunk}
\begin{CodeInput}
R> jvs_svy <- svydesign(ids = ~ 1, 
+                      weights = ~ weight,
+                      strata = ~ size + nace + region,
+                      data = jvs)
\end{CodeInput}
\end{CodeChunk}

Our second source (\code{admin}), the Central Job Offers Database, is a
register containing all vacancies submitted to Public Employment Offices
(see \url{https://oferty.praca.gov.pl}). We treat this register as a
non-probability sample since it contains administrative data provided on
a voluntary basis, so the inclusion mechanism is unknown. This dataset
was prepared in such a~way that records deemed to be out of scope
(either in terms of the definition of vacancy or the population of
entities) were excluded. In addition to the same variables found in the
JVS, the dataset contains one called \code{single\_shift}, which is our
target variable, defined as: whether a~company seeks at least one
employee for a~single-shift job. The goal of this case study is to
estimate the share of companies that seek employees for a~single-shift
job in Poland in a~given quarter.

\begin{CodeChunk}
\begin{CodeInput}
R> data("admin")
R> head(admin)
\end{CodeInput}
\begin{CodeOutput}
   id private size nace region single_shift
1 j_1       0    L    P     30        FALSE
2 j_2       0    L    O     14         TRUE
3 j_3       0    L    O     04         TRUE
4 j_4       0    L    O     24         TRUE
5 j_5       0    L    O     04         TRUE
6 j_6       1    L    C     28        FALSE
\end{CodeOutput}
\end{CodeChunk}

One should keep in mind that this paper does not aim to provide a
complete tutorial on how to use non-probability samples for statistical
inference. We therefore do not include the stage of aligning variables
to meet the same definitions, assessing the strength of the relation
between auxiliary variables and the target variable, the selection
mechanism and the distribution mis-matches between both samples. In the
examples below we assume that there is no overlap between both sources
and the naive, reference estimate, given by the mean of the
\texttt{single\_shift} column of \texttt{admin}, which is equal to
66.1\%.

\subsection{Estimation}\label{estimation}

\subsubsection{Propensity score
approach}\label{propensity-score-approach}

First, we start with the IPW approach, which offers the choice between
two estimation methods: MLE (default) and GEE (calibrated to the
estimated survey totals). We start by calling the \code{nonprob}
function, where we define the \code{selection} argument indicating which
variables are to be included, the \code{target} argument, which
specifies the variable of interest, i.e., \code{single_shift}. The
remaining arguments specify the \code{svydesign} object, the dataset and
the link function (\code{method_selection}; \code{"logit"} is the
default value). As the \code{pop\_size} argument was not specified, the
\(\hat{\mu}_{y,\text{IPW-H\'{a}jek}}\) is returned.

\begin{CodeChunk}
\begin{CodeInput}
R> ipw_est1 <- nonprob(
+   selection = ~ region + private + nace + size,
+   target = ~ single_shift,
+   svydesign = jvs_svy,
+   data = admin,
+   method_selection = "logit"
+ )
\end{CodeInput}
\end{CodeChunk}

In order to get the basic information about the estimated target
quantity we can use the \code{print} method to display the object. It
provides information about the methods, source of auxiliary variable,
variable selection, the naive (uncorrected estimator) and the corrected
along with the standard error and confidence interval. The the last
line, i.e., \code{selected estimator}, is limited to the maximum of 5
variables.

\begin{CodeChunk}
\begin{CodeInput}
R> ipw_est1
\end{CodeInput}
\begin{CodeOutput}
A nonprob object
 - estimator type: inverse probability weighting
 - method: logit (mle)
 - auxiliary variables source: survey
 - vars selection: false
 - variance estimator: analytic
 - population size fixed: false
 - naive (uncorrected) estimator: 0.6605
 - selected estimator: 0.7224 (se=0.0421, ci=(0.6399, 0.8048))
\end{CodeOutput}
\end{CodeChunk}

If we want to see detailed information about the approach, we can use
the \code{summary} method. This function provides information on the
sample sizes, sum of the IPW weights along with the distribution of IPW
weights (for non-probability sample) and propensity scores (denoted as
IPW probabilities) for both samples. A more advanced user may be
interested in inspecting the details of the fit by accessing the
\code{selection} element (i.e., \code{ipw_est1$selection}).

\begin{CodeChunk}
\begin{CodeInput}
R> summary(ipw_est1)
\end{CodeInput}
\begin{CodeOutput}
A nonprob_summary object
 - call: nonprob(data = admin, selection = ~region + private + nace + 
    size, target = ~single_shift, svydesign = jvs_svy, 
    method_selection = "logit")
 - estimator type: inverse probability weighting
 - nonprob sample size: 9344 (18%)
 - prob sample size: 6523 (12.6%)
 - population size: 51870 (fixed: false)
 - detailed information about models are stored in list element(s): 
    "selection"
----------------------------------------------------------------
 - sum of IPW weights: 52898.13 
 - distribution of IPW weights (nonprob sample):
   - min: 1.1693; mean: 5.6612; median: 4.3334; max: 49.9504
 - distribution of IPW probabilities (nonprob sample):
   - min: 0.0189; mean: 0.2894; median: 0.2525; max: 0.8552
 - distribution of IPW probabilities (prob sample):
   - min: 0.0200; mean: 0.2687; median: 0.2291; max: 0.8552
----------------------------------------------------------------
\end{CodeOutput}
\end{CodeChunk}

To extract the point estimate along with its standard error and 95\%
confidence interval in a~form of \code{data.frame} we can use the
\code{extract} method as shown below.

\begin{CodeChunk}
\begin{CodeInput}
R> extract(ipw_est1)
\end{CodeInput}
\begin{CodeOutput}
        target      mean         SE lower_bound upper_bound
1 single_shift 0.7223628 0.04207711   0.6398932   0.8048324
\end{CodeOutput}
\end{CodeChunk}

If we want to use the calibrated IPW approach, it is necessary to define
the \code{control_sel()} function in the \code{control_selection}
argument by setting the \code{est_method} argument equal to \code{"gee"}
(the default is \code{"mle"}) and the value of \code{gee_h_fun}.

\begin{CodeChunk}
\begin{CodeInput}
R> ipw_est2 <- nonprob(
+   selection = ~ region + private + nace + size,
+   target = ~ single_shift,
+   svydesign = jvs_svy,
+   data = admin,
+   method_selection = "logit",
+   control_selection = control_sel(gee_h_fun = 1, est_method = "gee")
+ )
\end{CodeInput}
\end{CodeChunk}

Results are comparable to the standard IPW point estimate (70.4 vs 72.2)
while the standard error is slightly higher.

\begin{CodeChunk}
\begin{CodeInput}
R> ipw_est2
\end{CodeInput}
\begin{CodeOutput}
A nonprob object
 - estimator type: inverse probability weighting
 - method: logit (gee)
 - auxiliary variables source: survey
 - vars selection: false
 - variance estimator: analytic
 - population size fixed: false
 - naive (uncorrected) estimator: 0.6605
 - selected estimator: 0.7042 (se=0.0398, ci=(0.6262, 0.7822))
\end{CodeOutput}
\end{CodeChunk}

The calibrated IPW significantly improves the balance, which can be
accessed via the \\ \code{check_balance()} function:

\begin{CodeChunk}
\begin{CodeInput}
R> data.frame(ipw_mle=check_balance(~size-1, ipw_est1, 1)$balance,
+            ipw_gee=check_balance(~size-1, ipw_est2, 1)$balance)
\end{CodeInput}
\begin{CodeOutput}
      ipw_mle ipw_gee
sizeL  -367.6       0
sizeM  -228.5       0
sizeS  1624.2       0
\end{CodeOutput}
\end{CodeChunk}

Notice that neither in the package nor in this paper do we provide a
detailed description of the post-hoc results, such as the covariate
balance. This can be done using existing CRAN packages, e.g., through
the \code{bal.tab()} function from the \pkg{cobalt} package
\citep{cobalt}.

\subsubsection{Prediction-based
approach}\label{prediction-based-approach}

If the user is interested in the prediction-based approach, in
particular involving MI estimators, then, they should specify the
argument \code{outcome} as a~formula (as in the case of the \code{glm()}
function). We allow a~single outcome (specified as
\code{y ~ x1 + x2 + ... + xk}) and multiple outcomes (as
\code{y1 + y2 + y3 ~ x1 + x2 + ... + xk}). Note that if the
\code{outcome} argument is specified, then there is no need to specify
the \texttt{target} argument. By default, the GLM type of an MI
estimator is used (i.e., \code{method_outcome = "glm"}). In the code
below we show how this type of an MI estimator can be declared.

\begin{CodeChunk}
\begin{CodeInput}
R> mi_est1 <- nonprob(
+   outcome = single_shift ~ region + private + nace + size,
+   svydesign = jvs_svy,
+   data = admin,
+   method_outcome = "glm",
+   family_outcome = "binomial"
+ )
R> 
R> mi_est1
\end{CodeInput}
\begin{CodeOutput}
A nonprob object
 - estimator type: mass imputation
 - method: glm (binomial)
 - auxiliary variables source: survey
 - vars selection: false
 - variance estimator: analytic
 - population size fixed: false
 - naive (uncorrected) estimator: 0.6605
 - selected estimator: 0.7032 (se=0.0112, ci=(0.6813, 0.7252))
\end{CodeOutput}
\end{CodeChunk}

In order to employ an MI estimator based on NN matching, one can specify
\code{method\_outcome = "nn"} for the nearest neighbours search using
all variables specified in the \code{outcome} argument,
\code{method\_outcome = "pmm"} to use PMM or
\code{method\_outcome = "npar"} to use non-parametric approach. For the
NN and PMM estimators, we employ \(k=5\) nearest neighbours (i.e.,
\code{control\_out(k=5)}) but the variance of the PMM estimator should
be estimated using bootstrap approach suggested by
\citet{chlebicki2025}. However, in this example we have decided not to
use bootstrap as all variables are categorical and this results in a
large number of NNs having the same distance and the numerical
approximation defined in \code{control\_out(eps)} may give slightly
different results between platforms. Therefore the reported variance is
based on the probability sample \(S_{\text{P}}\) only. In the case of
the MI-NN estimator there is no need to specify the
\code{family\_outcome} argument as no model is estimated underneath.

\begin{CodeChunk}
\begin{CodeInput}
R> mi_est2 <- nonprob(
+   outcome = single_shift ~ region + private + nace + size,
+   svydesign = jvs_svy,
+   data = admin,
+   method_outcome = "nn",
+   control_outcome = control_out(k = 5)
+ )
R> mi_est3 <- nonprob(
+   outcome = single_shift ~ region + private + nace + size,
+   svydesign = jvs_svy,
+   data = admin,
+   method_outcome = "pmm",
+   family_outcome = "binomial", 
+   control_outcome = control_out(k = 5)
+ )
\end{CodeInput}
\end{CodeChunk}

Results of both estimators are more or less similar, but it should be
noted that the NN version suffers from the curse of dimensionality, so
the PMM version seems to be more reliable.

\begin{CodeChunk}
\begin{CodeInput}
R> rbind("NN" = extract(mi_est2)[, 2:3], "PMM" = extract(mi_est3)[, 2:3])
\end{CodeInput}
\begin{CodeOutput}
         mean         SE
NN  0.6799537 0.01568503
PMM 0.7458724 0.01526712
\end{CodeOutput}
\end{CodeChunk}

As discussed in Section \ref{sec-methods}, IPW and MI estimators are
asymptotically unbiased only when the model and auxiliary variables are
correctly specified. To overcome this problem, the user can turn to
doubly robust estimators.

\subsubsection{The doubly robust
approach}\label{the-doubly-robust-approach}

In order to choose doubly robust estimation the user needs to specify
both the \texttt{selection} and \code{outcome} arguments. These formulas
can be specified with the same or varying number of auxiliary variables.
As in the MI approach, we also allow multiple outcomes. In the following
example code we have specified the non-calibrated IPW and the MI-GLM
estimator.

\begin{CodeChunk}
\begin{CodeInput}
R> dr_est1 <- nonprob(
+   selection = ~ region + private + nace + size,
+   outcome = single_shift ~ region + private + nace + size,
+   svydesign = jvs_svy,
+   data = admin,
+   method_selection = "logit",
+   method_outcome = "glm",
+   family_outcome = "binomial"
+ )
R> dr_est1
\end{CodeInput}
\begin{CodeOutput}
A nonprob object
 - estimator type: doubly robust
 - method: glm (binomial)
 - auxiliary variables source: survey
 - vars selection: false
 - variance estimator: analytic
 - population size fixed: false
 - naive (uncorrected) estimator: 0.6605
 - selected estimator: 0.7035 (se=0.0117, ci=(0.6806, 0.7263))
\end{CodeOutput}
\end{CodeChunk}

Detailed results can be displayed by using the \code{summary} method,
which prints both information about the distribution of the IPW weights
as well as the outcome residuals and predictions. For more details one
can access the \code{"outcome"} and \code{"selection"} fields in the
\code{nonprob} object.

\begin{CodeChunk}
\begin{CodeInput}
R> summary(dr_est1)
\end{CodeInput}
\begin{CodeOutput}
A nonprob_summary object
 - call: nonprob(data = admin, selection = ~region + private + nace + 
    size, outcome = single_shift ~ region + private + nace + 
    size, svydesign = jvs_svy, method_selection = "logit", 
    method_outcome = "glm", 
    family_outcome = "binomial")
 - estimator type: doubly robust
 - nonprob sample size: 9344 (18%)
 - prob sample size: 6523 (12.6%)
 - population size: 51870 (fixed: false)
 - detailed information about models are stored in list element(s): 
    "outcome" and "selection"
----------------------------------------------------------------
 - sum of IPW weights: 52898.13 
 - distribution of IPW weights (nonprob sample):
   - min: 1.1693; mean: 5.6612; median: 4.3334; max: 49.9504
 - distribution of IPW probabilities (nonprob sample):
   - min: 0.0189; mean: 0.2894; median: 0.2525; max: 0.8552
 - distribution of IPW probabilities (prob sample):
   - min: 0.0200; mean: 0.2687; median: 0.2291; max: 0.8552
----------------------------------------------------------------
 - distribution of outcome residuals:
   - single_shift: min: -0.9730; mean: 0.0000; median: 0.1075; max: 0.8564
 - distribution of outcome predictions (nonprob sample):
   - single_shift: min: 0.1436; mean: 0.6605; median: 0.6739; max: 0.9758
 - distribution of outcome predictions (prob sample):
   - single_shift: min: 0.1436; mean: 0.5930; median: 0.5938; max: 0.9785
----------------------------------------------------------------
\end{CodeOutput}
\end{CodeChunk}

Finally, we can use the bias minimization approach, as proposed by
\citet{yang_doubly_2020}, by specifying the
\code{control\_inference = control_inf(bias_correction = TRUE)} argument
together with the \code{vars\_combine = TRUE} and
\code{vars\_selection = TRUE} as this approach requires variable
selection followed by variable union from both equations.

\begin{CodeChunk}
\begin{CodeInput}
R> set.seed(2024)
R> dr_est2 <- nonprob(
+   selection = ~ region + private + nace + size,
+   outcome = single_shift ~ region + private + nace + size,
+   svydesign = jvs_svy,
+   data = admin,
+   method_selection = "logit",
+   method_outcome = "glm",
+   family_outcome = "binomial",
+   control_inference = control_inf(bias_correction = TRUE,
+                                   vars_combine = TRUE,
+                                   vars_selection = TRUE)
+ )
R> dr_est2
\end{CodeInput}
\begin{CodeOutput}
A nonprob object
 - estimator type: doubly robust
 - method: glm (binomial)
 - auxiliary variables source: survey
 - vars selection: true
 - variance estimator: analytic
 - population size fixed: false
 - naive (uncorrected) estimator: 0.6605
 - selected estimator: 0.7037 (se=0.0104, ci=(0.6833, 0.7240))
\end{CodeOutput}
\end{CodeChunk}

\subsection{Comparison of estimates}\label{comparison-of-estimates}

Finally, as there is no single method for non-probability samples, we
suggest comparing results in a~single table or a~plot. Figure
\ref{fig:comparison-of-est} presents point estimates along with 95\%
confidence intervals. The various estimators show interesting patterns
compared to the naive estimate (red dashed line). Both IPW estimators
are characterized with large standard errors and the point estimates
over 70\%. The MI estimators demonstrate notably different behaviours:
while MI-PMM produces the highest point estimate with the widest
confidence interval, MI-NN yields the lowest estimate, close to the
naive value. Results for the other estimators -- MI-GLM and DR (with and
without bias minimization) -- are clustered together, with similar point
estimates and confidence interval widths, suggesting some consensus in
their bias correction. All these methods indicate a~population parameter
higher than the naive estimate, but their relative consistency, provides
a certain degree of confidence in their bias correction capabilities.

\begin{CodeChunk}
\begin{CodeInput}
R> df_s <- rbind(extract(ipw_est1), extract(ipw_est2), extract(mi_est1),
+               extract(mi_est2), extract(mi_est3), extract(dr_est1), 
+               extract(dr_est2))
R> 
R> df_s$est <- c("IPW (MLE)", "IPW (GEE)", "MI (GLM)", "MI (NN)", 
+               "MI (PMM)", "DR", "DR (BM)")
R> 
R> ggplot(data = df_s, 
+        aes(y = est, x = mean, xmin = lower_bound, xmax = upper_bound)) + 
+   geom_point() + 
+   geom_vline(xintercept = mean(admin$single_shift), 
+              linetype = "dotted", color = "red") + 
+   geom_errorbar() + 
+   labs(x = "Point estimator and confidence interval", y = "Estimators") +
+   theme_bw()
\end{CodeInput}
\begin{figure}[ht!]

{\centering \includegraphics{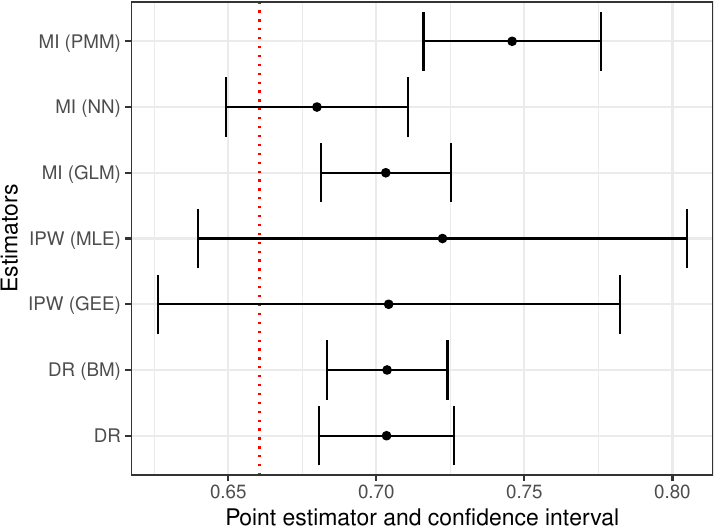} 

}

\caption[Comparison of estimates of the share of job vacancies offered on a~single-shift]{Comparison of estimates of the share of job vacancies offered on a~single-shift.}\label{fig:comparison-of-est}
\end{figure}
\end{CodeChunk}

\subsection{Advanced usage}\label{advanced-usage}

\subsubsection{Bootstrap approach for variance
estimation}\label{bootstrap-approach-for-variance-estimation}

In the package we allow the user to estimate the variance of the mean
analytically (by default) or using the bootstrap approach, as described
in Section \ref{sec-prediction}. We use analytical variance estimators
proposed in the papers referenced in Section \ref{sec-methods}. The
calculation of the standard error can be disabled using
\code{nonprob(se = FALSE)}. The bootstrap approach implemented in the
package refers to:

\begin{itemize}
\item the non-probability sample -- currently only simple random sampling with replacement is available via the \code{base::sample()},
\item the probability sample -- all the approaches implemented in the \code{as.svrepdesign()} function of the \pkg{survey} package are supported and we refer the reader to the relevant help file. 
\end{itemize}

The bootstrap approach is specified via the \code{control_inf()}
function with \code{var_method = "bootstrap"}. The bootstrap method for
the probability sample is controlled via the \code{rep_type} argument,
which passes the method to the \code{as.svrepdesign()} function. The
number of iterations is set in the \code{num_boot} argument (100 by
default). If the samples are large or the estimation method is
complicated (e.g., involves variable selection) one can set
\code{verbose = TRUE} to track progress. By default bootstrap results
are stored in the \code{boot_sample} element of the resulting list (to
disable this option, \code{keep_boot} should be set to \code{FALSE}).
The following code is an example of applying the IPW approach with the
bootstrap approach specified by the argument \code{control_inference} of
the \code{nonprob()} function.

\begin{CodeChunk}
\begin{CodeInput}
R> set.seed(2024)
R> ipw_est1_boot <- nonprob(
+   selection = ~ region + private + nace + size,
+   target = ~ single_shift,
+   svydesign = jvs_svy,
+   data = admin,
+   method_selection = "logit",
+   control_inference = control_inf(var_method = "bootstrap", num_boot = 50),
+   verbose = FALSE
+ )
\end{CodeInput}
\end{CodeChunk}

Next, we compare the estimated standard error of variance estimation for
the analytical and the bootstrap approach.

\begin{CodeChunk}
\begin{CodeInput}
R> rbind("IPW analytic variance"  = extract(ipw_est1)[, 2:3],
+       "IPW bootstrap variance" = extract(ipw_est1_boot)[, 2:3])
\end{CodeInput}
\begin{CodeOutput}
                            mean         SE
IPW analytic variance  0.7223628 0.04207711
IPW bootstrap variance 0.7223628 0.04307590
\end{CodeOutput}
\end{CodeChunk}

The boot samples can be accessed via the \code{boot_sample} element of
the output list of the \code{nonprob()} function. Note that the output
is returned as a~\code{matrix} because we allow multiple \code{target}
variables.

\begin{CodeChunk}
\begin{CodeInput}
R> head(ipw_est1_boot$boot_sample, n = 3)
\end{CodeInput}
\begin{CodeOutput}
     single_shift
[1,]    0.7406651
[2,]    0.6989083
[3,]    0.7667519
\end{CodeOutput}
\end{CodeChunk}

\subsubsection{Variable selection
algorithms}\label{variable-selection-algorithms}

In this section we briefly show how to use variable selection
algorithms. In order to indicate that a~variable selection algorithm
should be used one should specify the
\code{control_inference = control_inf(vars_selection = TRUE)} argument.
Then, the user should either leave the default setting or specify the
outcome parameters via the \code{control_out()} function or the
\code{control_sel()} function. Both functions have the same parameters:

\begin{itemize}
\item \code{penalty} -- The penalization function used during variables selection (possible values: \code{c("SCAD", "lasso", "MCP")}).
\item \code{nlambda} -- The number of $\lambda$ values; by default set to 50 (grid search).
\item \code{lambda_min} -- The smallest value for $\lambda$, as a~fraction of \code{lambda.max}; 0.001 by default.
\item \code{lambda} -- A user specified vector of lambdas (only for the \code{control_sel()} function).
\item \code{nfolds} -- The number of folds for cross validation; by default set to 10.
\item \code{a_SCAD, a_MCP} -- The tuning parameter of the SCAD and MCP penalty for the selection model; by default set to 3.7 and 3, respectively.
\end{itemize}

In the case of the MI approach we rely on the \pkg{ncvreg} package
\citep{ncvreg}, which is the only \proglang{R} package that employs the
SCAD method. For the IPW and DR approaches, we have developed our own
codes in \proglang{C++} via the \pkg{Rcpp} and \pkg{RcppArmadillo}
packages. In the code below we apply variable selection for the MI-GLM
estimator using only 5 folds, 25 possible values of \(\lambda\)
parameters and the LASSO penalty.

\begin{CodeChunk}
\begin{CodeInput}
R> set.seed(2024)
R> mi_est1_sel <- nonprob(
+   outcome = single_shift ~ region + private + nace + size,
+   svydesign = jvs_svy,
+   data = admin,
+   method_outcome = "glm",
+   family_outcome = "binomial" ,
+   control_outcome = control_out(nfolds = 5, nlambda = 25, penalty = "lasso"),
+   control_inference = control_inf(vars_selection = TRUE),
+   verbose = TRUE
+ )
\end{CodeInput}
\begin{CodeOutput}
Starting CV fold #1
Starting CV fold #2
Starting CV fold #3
Starting CV fold #4
Starting CV fold #5
\end{CodeOutput}
\end{CodeChunk}

In this case study, the MI-GLM estimator with variable selection yields
almost the same results as the approach without it. Point estimates and
standard errors differ at the fourth and third digit, respectively.

\begin{CodeChunk}
\begin{CodeInput}
R> rbind("MI without var sel" = extract(mi_est1)[, 2:3],
+       "MI with var sel"    = extract(mi_est1_sel)[, 2:3])
\end{CodeInput}
\begin{CodeOutput}
                        mean         SE
MI without var sel 0.7032089 0.01120237
MI with var sel    0.7019291 0.01102109
\end{CodeOutput}
\end{CodeChunk}

The result object of the \code{cv.ncvreg} class is stored in the
\code{"outcome"} element of the result, and you can access the selected
variables using the \code{coef} generic method as follows.

\begin{CodeChunk}
\begin{CodeInput}
R> round(coef(mi_est1_sel)$coef_out[, 1], 4)
\end{CodeInput}
\begin{CodeOutput}
(Intercept)    region04    region06    region08    region10    region12 
     0.2820      0.0025      0.3274      0.3196      0.2120      0.1775 
   region14    region16    region18    region20    region22    region24 
     0.0143      0.0792      0.0000      0.0000      0.0047     -0.2554 
   region26    region28    region30    region32     private     naceD.E 
     0.1333      0.0000      0.0000      0.0000     -0.6090      0.1759 
      naceF       naceG       naceH       naceI       naceJ     naceK.L 
     1.9173     -0.4558     -0.5607     -1.0966      0.9214      1.0370 
      naceM       naceN       naceO       naceP       naceQ     naceR.S 
     1.0025     -0.1840      1.4744      0.5368     -0.7116     -0.8138 
      sizeM       sizeS 
     0.9972      1.5354 
\end{CodeOutput}
\end{CodeChunk}

If a~user is interested in viewing the PS estimation results, then
access to the \code{"selection"} element is required.

\begin{CodeChunk}
\begin{CodeInput}
R> round(coef(ipw_est1)$coef_sel[, 1], 4)
\end{CodeInput}
\begin{CodeOutput}
(Intercept)    region04    region06    region08    region10    region12 
    -0.6528      0.8378      0.1995      0.1048     -0.1576     -0.6099 
   region14    region16    region18    region20    region22    region24 
    -0.8415      0.7639      1.1781      0.2225     -0.0375     -0.4067 
   region26    region28    region30    region32     private     naceD.E 
     0.2029      0.5786     -0.6102      0.3274      0.0590      0.7727 
      naceF       naceG       naceH       naceI       naceJ     naceK.L 
    -0.3778     -0.3337     -0.6517      0.4118     -1.4264      0.0617 
      naceM       naceN       naceO       naceP       naceQ     naceR.S 
    -0.4068      0.8003     -0.6935      1.2510      0.3029      0.2223 
      sizeM       sizeS 
    -0.3641     -1.0292 
\end{CodeOutput}
\end{CodeChunk}

\section[Classes and S3 methods]{Classes and \code{S3} methods}\label{sec-s3methods}

The package contains the main class \code{nonprob} and the supplementary
class \code{nonprob_method} and the related \code{nonprob_summary}
class. All available \code{S3} methods can be obtained by calling
\code{methods(class = "nonprob")}. For instance, the
\code{check_balance()} function, already mentioned in the case study, is
used to view the balance by checking how the PS weights reproduce known
or estimated population totals; the \code{nobs()} function returns the
sample size of the probability and non-probability samples.

\begin{CodeChunk}
\begin{CodeInput}
R> nobs(dr_est1)
\end{CodeInput}
\begin{CodeOutput}
   prob nonprob 
   6523    9344 
\end{CodeOutput}
\end{CodeChunk}

Table \ref{tab-s3methods} presents \code{S3} methods implemented for
objects of class \code{nonprob}. We have intentionally limited the
number of implemented methods as the goal of the package is to provide
point and interval estimates. If users are interested in assessing the
quality of the models or the covariate balance, they should use existing
\proglang{R} packages.

\begin{table}[ht!]
\centering
\begin{tabular}{p{3cm}p{12cm}}
\hline 
Function & Description \\
\hline
\code{check_balance} & returns a~\code{list} of totals for IPW and DR estimators \\
\code{coef} & returns a~list of coefficients for outcome or selection model \\
\code{confint} & returns the confidence interval for the target variable(s) \\
\code{extract} & returns a~\code{data.frame} with estimation results \\
\code{nobs} & returns the number of observations for both samples\\
\code{plot} & plots a~comparison of naive (uncorrected) and estimated mean along with CI\\
\code{pop_size} & returns the population size (fixed or not)\\
\code{summary} & returns a~\code{nonprob_summary} class with additional information\\
\code{update} & returns an \code{updated} object\\
\code{weights} & returns IPW weights for the IPW and DR estimator\\
\hline 
\end{tabular}
\caption{\code{S3} methods implemented for objects of class \code{nonprob} in \pkg{nonprobsvy}}
\label{tab-s3methods}
\end{table}

The confidence interval for the target variable(s) can be estimated
using the generic \code{confint()} function.

\begin{CodeChunk}
\begin{CodeInput}
R> confint(dr_est1, level = 0.99)
\end{CodeInput}
\begin{CodeOutput}
        target lower_bound upper_bound
1 single_shift   0.6734503   0.7334886
\end{CodeOutput}
\end{CodeChunk}

If a~user is interested in assessing the distribution of the IPW
weights, we suggest using the \code{weights()} function.

\begin{CodeChunk}
\begin{CodeInput}
R> summary(weights(dr_est1))
\end{CodeInput}
\begin{CodeOutput}
   Min. 1st Qu.  Median    Mean 3rd Qu.    Max. 
  1.169   2.673   4.333   5.661   7.178  49.950 
\end{CodeOutput}
\end{CodeChunk}

In the case of mass imputation estimators it is possible to use special
functions \code{method_glm()} that are of class \code{nonprob_method}.
An example is given below

\begin{CodeChunk}
\begin{CodeInput}
R> res_glm <- method_glm(
+   y_nons = admin$single_shift,
+   X_nons = model.matrix(~ region + private + nace + size, admin),
+   X_rand = model.matrix(~ region + private + nace + size, jvs),
+   svydesign = jvs_svy)
R> res_glm
\end{CodeInput}
\begin{CodeOutput}
Mass imputation model (GLM approach). Estimated mean: 0.7039 (se: 0.0115)
\end{CodeOutput}
\end{CodeChunk}

For the IPW we have created the function \code{method_ps()}, which
returns requested functions for a~specific link. This approach is
motivated by the following reasons: 1) we allow different estimation
techniques, such as maximum likelihood and general estimation equations;
2) the IPW estimator is also used for the DR estimator; and 3) variance
estimators depend on whether the IPW or the DR estimator is applied
(with combination with the MI estimator).

\begin{CodeChunk}
\begin{CodeInput}
R> method_ps()
\end{CodeInput}
\begin{CodeOutput}
Propensity score model with logit link
\end{CodeOutput}
\end{CodeChunk}

Details can be found in help page.

\begin{CodeChunk}
\begin{CodeInput}
R> ?method_ps()
\end{CodeInput}
\end{CodeChunk}

\section{Summary and future work}\label{summary-and-future-work}

The \pkg{nonprobsvy} package provides a~comprehensive \proglang{R}
software solution that addresses inference challenges connected with
non-probability samples by integrating them with probability samples or
known population totals/means. As non-probability data sources, like
administrative registers, voluntary online panels, and social media data
become increasingly available, statisticians need robust methods to
produce reliable population estimates. The package implements
state-of-the-art approaches including mass imputation, inverse
probability weighting, and doubly robust methods, each designed to
correct selection bias by leveraging auxiliary data. By providing a
unified framework and its integration with the \pkg{survey} package, the
\pkg{nonprobsvy} makes complex statistical methods for non-probability
samples more accessible, enabling researchers to produce robust
estimates even when working with non-representative data.

There are several avenues for future development of the \pkg{nonprobsvy}
package. One key priority is to implement model-based calibration and
additional methods for estimating propensity scores and weights. The
package currently assumes no overlap between probability and
non-probability samples, so accounting for potential overlap (e.g., in
big data sources and registers) is another important extension.
Additional planned developments include handling non-ignorable sample
selection mechanisms, developing a~theory for maintaining consistency
with calibration weights, and supporting multiple non-probability
samples from various sources for the purpose of data integration.
Further methodological extensions under consideration include empirical
likelihood approaches for doubly/multiply robust estimation, integration
of machine learning methods like debiased/double machine learning from
causal inference, handling measurement errors in big data variables, and
expanding the bootstrap approach beyond simple random sampling with
replacement.

The package will also be extended to handle the \texttt{svyrep.design}
class from the \pkg{survey} package and the \pkg{svrep} package. These
developments will enhance its capabilities for handling complex survey
data structures and modern estimation challenges.

\section{Acknowledgements}\label{sec-acknowledgements}

The authors' work has been financed by the National Science Centre in
Poland, OPUS 20, grant no. 2020/39/B/HS4/00941.

Łukasz Chrostowski was the main developer and maintainer of the package
up to version 0.1.0. Parts of this paper are based on Łukasz's Master's
thesis (available at
\url{https://github.com/ncn-foreigners/graduation-theses}). Piotr
Chlebicki has contributed to the package and has implemented MI-PMM
estimators. Maciej Beręsewicz was responsible for the design of the
package and for testing, reviewing and contributing to the source code.
He also prepared the manuscript. He was also responsible for a
significant restructuring of the package between versions 0.1.0 and
0.2.0.

The authors thank editors and reviewers for their constructive
suggestions.

\bibliography{references.bib}

\clearpage

\appendix

\section{Algorithms for the MI-NN and MI-PMM
estimators}\label{sec-details}

\begin{algorithm}[ht!]
\caption{Mass imputation using the $k$ nearest neighbour algorithm}
\label{algo-2}
\begin{algorithmic}[1]
\State If $k=1$, then for each $i \in S_{\text{P}}$ match $\hat{\nu}(i)$ such that
$\hat{\nu}(i)=
\operatornamewithlimits{arg\,min}_{j\in S_{\text{NP}}}d\left(\boldsymbol{x}_i,\boldsymbol{x}_j\right)$.
\State If $k>1$, then
$$\hat{\nu}(i, z) = \operatornamewithlimits{arg\,min}_{j\in S_{\text{NP}}\setminus\bigcup_{t=1}^{z-1}
\{\hat{\nu}(i, t)\}} d\left(\boldsymbol{x}_i, \boldsymbol{x}_j\right)$$
i.e., $\hat{\nu}(i, z)$ is $z$-th nearest neighbour from the sample $S_{\text{NP}}$.\;
\State For each $i \in S_{\text{P}}$, calculate the imputed value as

$$
y_i^* = \frac{1}{k}\sum_{t=1}^{k}y_{\hat{\nu}(i, t)}.
$$
\end{algorithmic}
\end{algorithm}

\begin{algorithm}[ht!]
\caption{Mass imputation using predictive mean matching variant: $\hat{y}-\hat{y}$ matching}
\label{algo-3}
\begin{algorithmic}[1]
\State Estimate regression model $m(\boldsymbol{x}, \boldsymbol{\beta})$ parameters.\;
\State Predict 

$$\hat{y}_{i}=m\left(\boldsymbol{x}_{i},\hat{\boldsymbol{\beta}}\right),  \hat{y}_{j}=m\left(\boldsymbol{x}_{j},\hat{\boldsymbol{\beta}}\right)$$

for $i\in S_{\text{P}}, j\in S_{\text{NP}}$ and assign each  $i\in S_{\text{P}}$ to $\hat{\nu}(i)$, where

$$
\hat{\nu}(i)=
\operatornamewithlimits{arg\,min}_{j\in S_{\text{NP}}}d\left(\hat{y}_{i},\hat{y}_{j}\right).\;
$$ 

\State If $k>1$, then:
$$
\hat{\nu}(i, z) = \operatornamewithlimits{arg\,min}_{ j\in S_{\text{NP}}\setminus\bigcup_{t=1}^{z-1}
\{\hat{\nu}(i, t)\}} d\left(\hat{y}_{i},\hat{y}_{j}\right)
$$
e.g., $\hat{\nu}(i, z)$ is $z$-th nearest neighbour from a~sample $S_{\text{NP}}$.\;
\State For $i \in S_{\text{P}}$, calculate imputation value as 
$$
y_i^* = \frac{1}{k}\sum_{t=1}^{k}y_{\hat{\nu}(i, t)}.
$$
\end{algorithmic}
\end{algorithm}

\begin{algorithm}[ht!]
\caption{Mass imputation using predictive mean matching variant: $\hat{y}-y$ matching}
\label{algo-4}
\begin{algorithmic}[1]

\State Estimate regression model $m(\boldsymbol{x}, \boldsymbol{\beta})$ parameters.\;

\State Predict 
$$
\hat{y}_{i}=m\left(\boldsymbol{x}_{i},\hat{\boldsymbol{\beta}}\right)
$$  for $i \in S_{\text{P}}$ 
and assign each  $i \in S_{\text{P}}$ to $\hat{\nu}(i)$, where
$$
\hat{\nu}(i)=
\operatornamewithlimits{arg\,min}_{j \in S_{\text{NP}}}d\left(\hat{y}_{i},y_{j}\right).\;
$$
\State If $k>1$, then:
$$
\hat{\nu}(i, z) = \operatornamewithlimits{arg\,min}_{j \in S_{\text{NP}} \setminus \bigcup_{t=1}^{z-1}
\{\hat{\nu}(i, t)\}} d\left(\hat{y}_{i},y_{j}\right).
$$
\State For each $i \in S_{\text{P}}$ calculate imputation value as
$$
y_i^* = \frac{1}{k}\sum_{t=1}^{k}y_{\hat{\nu}(i, t)}.
$$
\end{algorithmic}
\end{algorithm}

\clearpage

\end{document}